\documentclass[aps,prapplied,
    letterpaper,
    amsfonts,
    amsmath,
    reprint,
    superscriptaddress,
    longbibliography,
    nofootinbib,
    floatfix
]{revtex4-2}
\usepackage[english]{babel}
\usepackage[utf8]{inputenc}
\usepackage{xcolor}
\usepackage{svg}
\usepackage{graphicx}

\newcommand{\kett}[1]{|\,#1\rangle\!\rangle}
\newcommand{\bbra}[1]{\langle\!\langle#1\,|}
\newcommand{\bbrakett}[2]{\langle\!\langle#1\,|\,#2\rangle\!\rangle}
\newcommand{\ket}[1]{|#1\rangle}
\newcommand{\bra}[1]{\langle#1|}
\usepackage{silence}
\usepackage[hidelinks]{hyperref}
\hypersetup{
  pdftitle={Dynamically suppressing cavity dephasing induced by frequency fluctuations of a coupled nonlinear mode},
  pdfauthor={Yunwei Lu; Xinyuan You; Ziwen Huang; S\'ebastien L\'eger; Jens Koch; Yao Lu},
  pdfsubject={Stark-Assisted Flux-noise Evasion in superconducting cavities},
  pdfkeywords={bosonic quantum information, cavity dephasing, flux noise, dynamical sweet spot, superconducting circuits}
}

\begin{document}
\title{Dynamically suppressing cavity dephasing induced by frequency fluctuations of a coupled nonlinear mode}

\author{Yunwei Lu}
\email{yunweilu2020@u.northwestern.edu}
\affiliation{Department of Physics and Astronomy,
Northwestern University, Evanston, Illinois 60208, USA}

\author{Xinyuan You}
\affiliation{Fermi National Accelerator Laboratory,
Batavia, Illinois 60510, USA}

\author{Ziwen Huang}
\altaffiliation[Current address: ]{Amazon Center for Quantum Computing}
\affiliation{Fermi National Accelerator Laboratory,
Batavia, Illinois 60510, USA}

\author{S\'ebastien L\'eger}
\affiliation{Departments of Physics and Applied Physics,
Stanford University, Stanford, California 94305, USA}

\author{Jens Koch}
\email{jens-koch@northwestern.edu}
\affiliation{Department of Physics and Astronomy,
Northwestern University, Evanston, Illinois 60208, USA}

\author{Yao Lu}
\email{yaolu@fnal.gov}
\affiliation{Fermi National Accelerator Laboratory,
Batavia, Illinois 60510, USA}

\begin{abstract}
High-coherence superconducting cavities offer a promising platform for quantum information, with long coherence times and negligible intrinsic dephasing. However, cavity control generally relies on nonlinear Josephson elements whose frequency fluctuations are inherited by the cavity as dephasing, potentially limiting control fidelities and eroding the noise bias used in error-correction protocols. Here, we introduce a hardware-efficient technique, called Stark-Assisted Flux-noise Evasion (SAFE), for suppressing inherited cavity dephasing using only a weak, off-resonant microwave drive applied to the nonlinear element. As a concrete setup, we analyze a 3D superconducting cavity dispersively coupled to a flux-tunable transmon (FTT) subject to $1/f$ flux noise. Analytical predictions are confirmed by Monte Carlo simulations with realistic parameters, which show that SAFE can extend the cavity dephasing time by more than an order of magnitude while keeping residual drive-induced decoherence subdominant.
\end{abstract}

\maketitle

\section{Introduction}

Superconducting microwave cavities are a promising platform for quantum information processing because they exhibit long coherence times \cite{longcoherence1, longcoherence2, longcoherence3, longcoherence4, longcoherence5}. This advantage arises in part because cavities avoid Josephson-junction-related noise mechanisms, including $1/f$ critical-current noise~\cite{Wellstood2004}, and junction aging~\cite{Gates1984}. These properties make cavities attractive as quantum memories and logical qubits~\cite{Krasnok2024CavitiesQubits}. Hardware-efficient bosonic quantum error-correction schemes further strengthen this platform by protecting encoded information against dominant errors~\cite{breakeven1,breakeven2,breakeven3,breakeven4,breakeven5,breakeven7,aqec1,aqec2,Cai2021BosonicQEC}. In particular, cat and dual-rail encodings exploit the strong bias toward photon loss over dephasing in cavities, which underlies their error-correction advantage~\cite{teoh2023dual,putterman2025hardware}.


Universal control of bosonic qubits generally requires coupling the cavity
to a Josephson ancillary or coupling element
~\cite{Ma2021QuantumControl,cavitycontrol1,cavitycontrol2,cavitycontrol3}.
The same hybridization that provides state preparation, tunable interactions,
and readout also opens channels through which the cavity can inherit relaxation
and dephasing from the ancillary circuit
~\cite{SNAIL,magnetichose1,serge,Ding2024KerrCat}.
Flux-tunable Josephson elements, such as nonlinear ancillas and parametric
couplers, provide particularly versatile control over individual bosonic modes
and interactions among them
~\cite{SNAIL,squid,linc,Atanasova2025Fluxonium,li2025,
valadares2026,Copetudo2026DirectPhase}.
However, the flux tunability often comes with low-frequency flux noise, resulting in cavity dephasing that can directly compete with the favorable photon-loss bias exploited by
bosonic error-correction schemes. Related work has also explored suppressing cavity frequency noise using a Kerr nonlinearity~\cite{vanSoest2026Suppressing}. A common strategy to mitigate this dephasing is to idle the nonlinear element at a DC flux sweet spot\ \cite{Neill2018, Barends2014} that is insensitive to flux noise, and tune it away from the sweet spot only when an operation requires it. However, this scheme is challenging in two respects: (1) in three-dimensional cavities, ensuring both fast flux tuning and a high-coherence cavity requires extra hardware complexity~\cite{magnetichose1,magnetichose2,li2025,valadares2026}; (2) tuning flux carries the risk of inducing unwanted transitions~\cite{Sung2021CZiSWAPleakage,Yan2018TunableCouplingleakage,Stehlik2021TunableCouplingleakage}.

Here, based on the concept of dynamical sweet-spot engineering in driven superconducting qubits~\cite{PhysRevApplied.12.054015dss2,Hong2020ProtectedGate,Mundada2020FloquetFluxonium,PhysRevApplied.15.034065floquetqubit,Valery2022TwoTone,Gandon2022FloquetQubits,Cheng2022ComplexAmplitude,Thibodeau2024FloquetMolecule,BrisenoColunga2026SweetSour,Lauwens2026TwoTone,Yang2026ParetoDSS}, we introduce an alternative dynamical approach that leaves the DC flux bias unchanged. By applying a weak off-resonant drive to the nonlinear circuit, we induce an ac Stark shift that can be engineered to cancel the leading flux sensitivity inherited by the cavity. We therefore call this approach Stark-Assisted Flux-noise Evasion (SAFE). With suitable drive parameters, the shifted cavity frequency is first-order insensitive to flux noise, thereby creating a dynamical sweet spot. For a cavity coupled to a flux-tunable transmon with a representative set of parameters, Monte Carlo simulations show an approximately \(20\)-fold improvement in the cavity dephasing time.

The paper is organized as follows. Section~\ref{sec2} introduces the mechanism of SAFE for suppressing cavity dephasing using a drive and shows how to select the corresponding drive parameters. In Section~\ref{sec3}, we derive the drive-induced dephasing rates and compare them with numerical simulations. In Section~\ref{sec4}, we apply the SAFE scheme to binomial and dual-rail encodings, highlighting the ability of SAFE to suppress dephasing simultaneously across multiple pairs of Fock states and multiple modes. Section~\ref{sec5} summarizes our main results.

\section{Dynamical Suppression of Cavity Dephasing}  
\label{sec2}
This section develops the basic SAFE mechanism in a minimal setting: a cavity coupled to a nonlinear device, specifically, a flux-tunable transmon (FTT).
Although we focus on the FTT--cavity system for concreteness, SAFE can be applied to cavities coupled to other nonlinear superconducting qubits or couplers.


\subsection{Cavity Dephasing from Flux Noise}
\label{subsec:setup}

\begin{figure*}[t]
\centering
\includegraphics[width=0.8\textwidth]{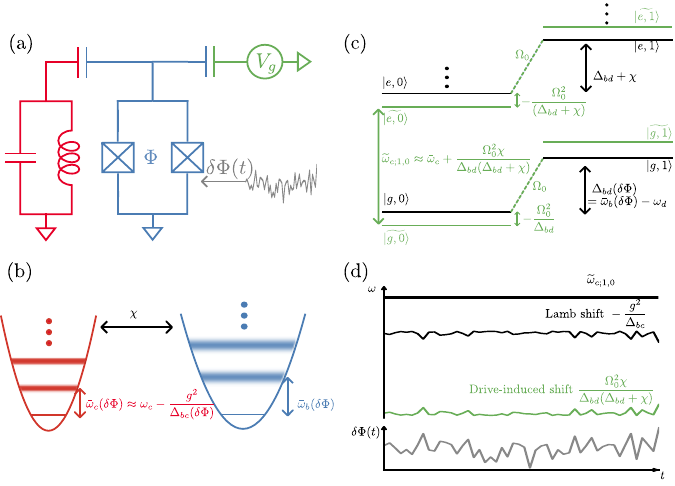}
\caption{
Illustration of suppressing cavity dephasing by Stark-assisted flux-noise evasion (SAFE).
\textbf{(a)} Circuit diagram of a flux-tunable transmon (FTT) dispersively coupled to a superconducting cavity. The FTT is driven by a coherent voltage source $V_g$. Flux noise $\delta\Phi(t)$ modulates the FTT frequency and constitutes the dominant noise source.
\textbf{(b)} Fluctuations of the cavity frequency induced by the Lamb shift. The bare modes have frequencies $\omega_c$ and $\omega_b(\delta\Phi)$, with detuning $\Delta_{bc}(\delta\Phi)=\omega_b(\delta\Phi)-\omega_c$. The coupling $g$ hybridizes the modes and yields dressed frequencies $\bar{\omega}_{b/c}(\delta\Phi)$, dispersive shift $\chi$. As a result, fluctuations in the FTT frequency $\delta\omega_b(t)$ caused by flux noise $\delta\Phi(t)$ lead to cavity-frequency fluctuations $\delta\bar{\omega}_c(t)$ via the Lamb shift $-g^2/\Delta_{bc}$.
\textbf{(c)} Drive-induced AC Stark shift in the frame rotating at the drive frequency $\omega_d$. For a fixed cavity photon number $m$, the drive hybridizes the states $\ket{g,m}$ and $\ket{e,m}$, generating level repulsion. Because the relevant FTT--drive detuning $\Delta_{bd}+m\chi$ is dispersively shifted by the cavity photon number, the AC Stark shift depends on $m$ and therefore contributes to the shifted cavity transition frequency $\widetilde{\omega}_{c;1,0}$.
\textbf{(d)} Dynamical sweet spot. By tuning the drive-induced AC Stark shift (green), the shifted cavity transition frequency between neighboring Fock states (thick black line) can be made first-order insensitive to flux noise $\delta\Phi(t)$ (gray), thereby compensating the static Lamb-shift contribution (black).
}
\label{fig:device}
\end{figure*}

Figure~\ref{fig:device}(a) illustrates the system considered here: a flux-tunable transmon (FTT) is capacitively coupled to a cavity and driven by an external microwave field. The full Hamiltonian describing this system is
\begin{align}
\label{Hfull}
&H_{\text{full}}\!\left(t;\Phi+\delta\Phi(t)\right)
=
\nonumber \\
&4E_C (n-n_g)^2
-E_{J}\!\left(\Phi+\delta\Phi(t)\right)
\cos\!\left[
\varphi-\varphi_0\!\left(\Phi+\delta\Phi(t)\right)
\right]
\nonumber \\
&\quad +
\omega_c c^{\dagger}c
+
g' n (c^{\dagger}+c)
+
2\Omega'_0\cos(\omega_d t) n.
\end{align}
Throughout this work, we set $\hbar=1$. Here, $E_C$ is the charging energy, $n$ and $\varphi$ are the charge and phase operators of the FTT, and $n_g$ is the offset charge. The effective Josephson energy $E_{J}(\Phi+\delta\Phi)$ and the potential phase offset $\varphi_0(\Phi+\delta\Phi)$ depend on the external flux $\Phi$ and the flux fluctuation $\delta\Phi(t)$. The offset describes the flux-dependent shift of the Josephson-potential minimum caused by asymmetry between the two junctions in the SQUID loop. The operator $c$ annihilates a photon in the cavity mode with frequency $\omega_c$. The capacitive coupling strength is $g'$, and the last term describes a charge drive applied to the FTT, with amplitude $\Omega'_0$ and frequency $\omega_d$.

For small flux fluctuations \(\delta\Phi(t)\ll\Phi\), expanding \(H_{\text{full}}\) to first order in $\delta\Phi(t)$ gives
\begin{equation}
H_{\text{full}}\!\left(t;\Phi+\delta\Phi(t)\right)
\approx H_{\text{sys}}(\Phi) + H_{\text{drive}}(t)+\underbrace{\delta\Phi(t)\frac{\partial H_{\text{sys}}(\Phi)}{\partial \Phi}}_{H_{\text{noise}}},
\end{equation}
where the system, drive, and noise Hamiltonians are 
\begin{equation}
\begin{aligned}
\label{system hamiltonian}
H_{\text{sys}}(\Phi) =&\;  4E_C (n-n_g)^2
- E_{J}(\Phi)
\cos\!\left[{\varphi}-{\varphi_0(\Phi)}\right] \\
&+ \omega_c c^{\dagger}c
+ g' n (c^{\dagger}+c)\\
H_{\text{drive}}(t) = &\,2\Omega'_0\cos(\omega_d t)\,n\\
H_{\text{noise}}(t) =&\, -\delta\Phi(t)\biggl[ 
\frac{\partial E_{J}(\Phi)}{\partial\Phi} \cos(\varphi - \varphi_0(\Phi)) \\
&+ E_{J}(\Phi) \frac{\partial \varphi_0(\Phi)}{\partial\Phi} \sin(\varphi - \varphi_0(\Phi)) 
\biggr].
\end{aligned}
\end{equation}
For compactness, we do not show the explicit \(\Phi\) dependence of
\(E_{J}\), \(\varphi_0\), and $H_{\text{sys}}$ below.

In the transmon regime ($E_C \ll E_{J}$), the wave function is localized near a potential minimum, and the phase and charge operators can be expressed in terms of ladder operators as
\begin{equation}
\begin{aligned}
\varphi - \varphi_0 &= \varphi_{\text{ZPF}} \left( b + b^\dagger \right), \quad
n = \frac{-i \left( b - b^\dagger \right)}{2 \varphi_{\text{ZPF}}},
\end{aligned}
\end{equation}
with $\varphi_{\text{ZPF}} = \left( {2 E_C}/{E_{J}} \right)^{1/4}$.
Expanding the cosine potential to fourth order in $\varphi-\varphi_0$ and applying the rotating-wave approximation (RWA) yields
\begin{equation}
\begin{split}
\label{sysHamiltonian}
H_\text{sys} &\approx \omega_{b} b^{\dagger} b
+ \frac{K}{2} b^{\dagger 2} b^{2}
+ \omega_c c^\dagger c
+ g ( b + b^{\dagger})( c + c^{\dagger}) .
\end{split}
\end{equation}
The FTT frequency and anharmonicity are $\omega_{b} = \sqrt{8E_{C}E_{J}} - E_{C}$ and $K = -E_{C}$, respectively. The rescaled drive amplitude and coupling strength are $\Omega_0 = \Omega'_0/(2\varphi_{\text{ZPF}})$ and $g = g'/(2\varphi_{\text{ZPF}})$, respectively.

We expand the noise Hamiltonian $H_{\text{noise}}(t)$, keeping terms through second order in $\varphi_{\mathrm{ZPF}}$:
\begin{equation}
\begin{aligned}
H_{\text{noise}}(t) &\approx
-\delta\Phi(t)\biggl[
\frac{\partial E_{J}}{\partial\Phi}
\left(1-\frac{\varphi_{\mathrm{ZPF}}^2}{2}(b+b^\dagger)^2\right) \\
&\qquad\qquad + E_{J}\frac{\partial \varphi_0}{\partial\Phi}
\varphi_{\mathrm{ZPF}}(b+b^\dagger)
\biggr].
\end{aligned}
\end{equation}
This second-order approximation is sufficient to capture the depolarization and pure-dephasing channels induced by flux noise, of which the latter will show dominant. The terms proportional to $b+b^\dagger$ and $b^2+b^{\dagger2}$ allow the fluctuating flux to induce stochastic transitions at angular frequencies $\omega_b$ and $2\omega_b$ while the number-conserving term $2b^\dagger b$ leads to pure dephasing. For $1/f$ flux noise with spectral density
\begin{equation}
\label{1/fdefine}
S_{1/f}(\omega)
= \int_{-\infty}^{\infty} d\tau \, \langle \delta\Phi(\tau) \delta\Phi(0)\rangle e^{i\omega \tau}
= \frac{A^2}{\left|\omega/2\pi\right|},
\end{equation}
the corresponding depolarization rates are governed by $S_{1/f}(\omega_b)$ and $S_{1/f}(2\omega_b)$. Because $S_{1/f}(\omega)$ is strongly suppressed at the relevant transition frequencies ($\omega_b/2\pi\sim\mathrm{GHz}$), the depolarization rates are negligible. On the other hand, the pure-dephasing rate induced by the number-conserving term is  governed by the low-frequency noise spectrum $S_{1/f}(\omega\to0)$. Retaining only the number-conserving contribution from the quadratic term gives
\begin{equation}
H_{\text{noise}}(t)\approx
\frac{\partial E_{J}}{\partial\Phi} \varphi_{\mathrm{ZPF}}^2
\delta\Phi(t) b^\dagger b .
\end{equation}
Using
\begin{equation}
\frac{\partial\omega_b}{\partial\Phi}
= \frac{\partial\omega_b}{\partial E_{J}}\frac{\partial E_{J}}{\partial\Phi}
= \varphi_{\mathrm{ZPF}}^2 \frac{\partial E_{J}}{\partial\Phi},
\label{eq:sensitivity_deriv}
\end{equation}
one obtains
\begin{equation}
\label{noisyhamiltonian}
H_{\text{noise}}(t)\approx
\frac{\partial\omega_b}{\partial\Phi} \delta\Phi(t) b^\dagger b .
\end{equation}
Consequently, the primary effect of low-frequency flux noise is random modulation of the FTT transition frequency.

In superconducting implementations of bosonic qubits, a transmon ancilla is typically dispersively coupled to the storage cavity for control and readout \cite{breakeven1,Heeres2017UniversalGateSetdispersive,eickbusch2022ecd}. Specifically, we assume \(g/|\Delta_{bc}|\ll1\) and \(g/|\Delta_{bc}+K|\ll1\), where \(\Delta_{bc}\equiv\omega_b-\omega_c\). In this regime, the system Hamiltonian \(H_{\mathrm{sys}}\) is well approximated by
\cite{cqedreview2021} 
\begin{multline}
\label{Hdisp}
H_\text{disp} =
\bar{\omega}_b\, b^\dagger b
+ \frac{K}{2}\, b^{\dagger 2} b^{2}
+ \bar{\omega}_c\, c^\dagger c
+ \chi\, b^\dagger b\, c^\dagger c .
\end{multline}
Here, \(\chi\) is the dispersive shift, while \(\bar{\omega}_b\) and \(\bar{\omega}_c\) are the Lamb-shifted FTT and cavity frequencies:
\begin{align}
\label{dispersive parameters}
\chi &\approx 2\frac{g^2}{\Delta_{bc}^2}K,\quad
\bar{\omega}_b \approx \omega_b + \frac{g^2}{\Delta_{bc}}, \quad
\bar{\omega}_c \approx \omega_c - \frac{g^2}{\Delta_{bc}},
\end{align}
where we expand \(\chi\) to first order in \(K/\Delta_{bc}\).
In the dispersive frame, $H_{\mathrm{noise}}(t)$ contains both longitudinal and transverse terms, giving rise to pure dephasing and depolarization, respectively. Typically, the latter oscillate at frequencies in the GHz range, and therefore probe the $1/f$ noise spectrum at correspondingly high frequencies. Since the spectral weight of $1/f$ noise is negligible at these frequencies, these transverse terms can be safely neglected. Consequently, $H_{\mathrm{noise}}(t)$ primarily gives rise to pure dephasing in the dispersive frame. In particular, fluctuations in the FTT frequency \(\delta\omega_b(t)\) lead to fluctuations in the cavity frequency through the Lamb shift
\begin{align}
\delta\bar{\omega}_c(t)
= \frac{\partial \bar{\omega}_c}{\partial \omega_b}\,\delta\omega_b(t)
\approx \frac{g^2}{\Delta_{bc}^2}\,\delta\omega_b(t),
\end{align}
as shown in Fig.~\ref{fig:device}(b). Such fluctuations produce cavity dephasing.
\subsection{Dynamical Sweet Spots Generated by SAFE}
\label{subsec:ssconditiondiscussion}
\begin{figure}[t]
    \centering
    \includegraphics[width=0.8\linewidth]{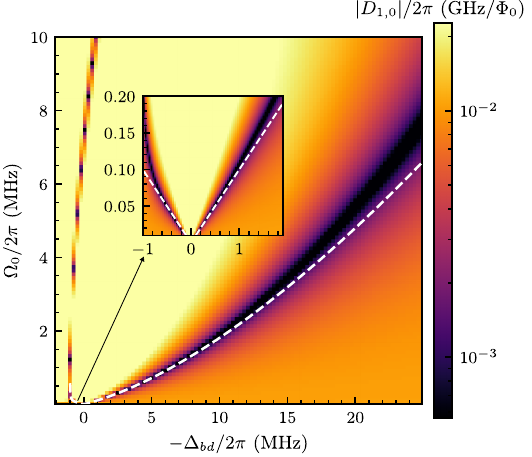}
    \caption{Dynamical sweet spots generated by SAFE. Two-dimensional map of \(D_{1,0}\) as a function of drive amplitude \(\Omega_0\) and detuning \(\Delta_{bd}\). Dashed curves denote the analytical sweet-spot conditions from Eqs.~\eqref{eq:ss_unselective} and \eqref{ss_selective} for the main panel and the inset, respectively. See App.~\ref{app:numerics} for the representative system parameters in the dispersive regime.}
\label{fig:compare}
\end{figure}

Previous work has shown that qubits can be protected from $1/f$ noise by engineering dynamical sweet spots~\cite{PhysRevApplied.15.034065floquetqubit}. The central idea is to apply a continuous periodic drive to the system. For a periodic Hamiltonian, $H(t+T)=H(t)$, the Floquet modes $\ket{\psi_\alpha(t)}$ satisfy
\begin{equation}
\bigl(H(t)-i\partial_{t}\bigr) \ket{\psi_\alpha(t)} = \varepsilon_\alpha\ket{\psi_\alpha(t)}.
\end{equation}
where $\varepsilon_\alpha$ are the Floquet quasienergies. The transition frequencies between Floquet states are determined by the corresponding quasienergy differences. A dynamical sweet spot is an operating point at which a continuous periodic drive renders these transition frequencies first-order insensitive to low-frequency noise fluctuations. Here, we apply the same principle to the FTT--cavity system by considering a general cavity transition between Floquet states adiabatically connected to $\ket{g,n}$ and $\ket{g,m}$, with $n$ and $m$ arbitrary cavity photon numbers. The corresponding dynamical sweet-spot condition is
\begin{equation}
\frac{\partial \omega^{(\mathrm F)}_{c;n,m}}{\partial \Phi}=0,
\end{equation}
where $\omega^{(\mathrm F)}_{c;n,m}$ is the corresponding Floquet cavity transition frequency.

In contrast to the weak-drive dynamical sweet spots of Ref.~\cite{PhysRevApplied.15.034065floquetqubit}, which emerge near multiphoton resonances, the sweet spots identified here occur in a weak, off-resonant driving regime. In this regime, the quasienergies and Floquet states of the time-periodic Hamiltonian $H_\text{sys}+H_{\text{drive}}(t)$ [Eq.\ \eqref{system hamiltonian}] can be approximated using a time-independent Hamiltonian in a frame rotating at the drive frequency. Transforming $H_{\text{disp}}+H_{\text{drive}}(t)$ into the rotating frame $U_R(t)=e^{-i\omega_d t b^\dagger b}$ and applying the rotating-wave approximation, we obtain
\begin{equation}
\label{eq:H_R_floquet}
H_R = \Delta_{bd} b^\dagger b + \frac{K}{2} b^{\dagger 2} b^{2} + \overline{\omega}_c c^\dagger c + \chi b^\dagger b c^\dagger c + \Omega_0 ( b + b^\dagger ),
\end{equation}
where $\Delta_{bd}=\bar{\omega}_b-\omega_d$ is the FTT--drive detuning. The transformation of the noise Hamiltonian is discussed later in Sec.~\ref{sec3}. The eigenvalues of $H_R$ approximate the Floquet quasienergies, up to integer multiples of the drive frequency $\omega_d$. The corresponding laboratory-frame Floquet states are approximated by applying $U^\dagger(t)=e^{i\omega_d t b^\dagger b}$ to the eigenstates of $H_R$, as justified in App.~\ref{app:floquet}.

An analytical expression for the eigenvalues of $H_R$ can be derived in the weak-drive regime by treating the drive term perturbatively. Among these
eigenvalues, we focus on the cavity transition frequencies with the
FTT in its ground state. Keeping the leading correction in
$\Omega_0$, we obtain the shifted cavity transition frequency
\begin{align}
\label{energy correction}
\widetilde{\omega}_{c;n,m}=
(n-m)
\left(
\omega_c-\frac{g^2}{\Delta_{bc}}
\right)
- \frac{\Omega_0^2}{\Delta_{bd}+n\chi}
+ \frac{\Omega_0^2}{\Delta_{bd}+m\chi}.
\end{align}
The last two terms arise from the drive-induced level repulsion of the states
$\ket{g,n}$ and $\ket{g,m}$. For a fixed cavity photon number $m$, the
drive couples $\ket{g,m}$ to states with an excited FTT and lowers its energy by an
amount proportional to
$-\Omega_0^2/(\Delta_{bd}+m\chi)$. The denominator is the shifted
FTT--drive detuning,
$\Delta_m\equiv\Delta_{bd}+m\chi$, where the $m\chi$ term arises from the
dispersive shift of the FTT transition. As a result, different cavity Fock
states acquire different drive-induced energy shifts, as shown in
Fig.~\ref{fig:device}(c).
Because these shifts depend on the FTT frequency, they also modify the flux
sensitivity of the cavity transition frequencies. In the presence of flux
noise, the shifted cavity transition frequency fluctuates as
\begin{equation}
\delta\widetilde{\omega}_{c;n,m}(t)
= \frac{\partial \widetilde{\omega}_{c;n,m}}
{\partial \Phi}
\delta\Phi(t).
\end{equation}

We characterize the sensitivity of the shifted cavity transition frequency to
flux fluctuations by
\begin{equation}
D_{n,m}
= \frac{\partial \widetilde{\omega}_{c;n,m}}{\partial \Phi}.
\end{equation}
By tuning the drive parameters, we can generate a sweet spot at which the
cavity transition frequency is first-order insensitive to flux noise, i.e., $D_{n,m}=0$, as illustrated in Fig.~\ref{fig:device}(d). Depending on the drive detuning
$\Delta_{bd}$, a sweet spot can be realized either for a specific cavity
transition or for multiple cavity transitions.

\paragraph{Large-detuning regime.}
\label{large-detuning regime}
In this case, the drive detuning $\Delta_{bd}$ satisfies $|\Delta_{bd}|\gg |\chi|$. Expanding Eq.~\eqref{energy correction} to first order in $\chi/\Delta_{bd}$ gives the shifted cavity transition frequency
\begin{equation}
\widetilde{\omega}_{c;n,m}\approx(n-m)\bigg({\omega}_c-\frac{g^2}{\Delta_{bc}}+\frac{\Omega_0^2\chi}{\Delta_{bd}^2}\bigg).
\end{equation}
The corresponding susceptibility is
\begin{equation}
\label{eq:D01_unselective}
D_{n,m}
\approx
(n-m) \frac{g^2}{\Delta_{bc}^2}
\bigg(
1 - \frac{4\Omega_{0}^{2} }{\Delta_{bd}^2} \frac{K}{\Delta_{bd}}
\bigg)
\frac{\partial\omega_b}{\partial \Phi}.
\end{equation}
Choosing the detuning
\begin{equation}
\label{eq:ss_unselective}
\Delta_{bd}=(4\Omega_0^2K)^{1/3}
\end{equation}
generates a sweet spot. Since the FTT has negative anharmonicity ($K<0$), the sweet spot occurs at negative detuning. At the sweet spot, the weak-drive condition imposes the additional constraint
\begin{equation}
\frac{\Omega_0}{|\Delta_{bd}|}
= \sqrt{\frac{\Delta_{bd}}{4K}} \ll 1
\Rightarrow |\Delta_{bd}| \ll |K|.
\end{equation}
The leading-order expression above is independent of the cavity photon number apart from the factor $(n-m)$. This suggests that the same choice of detuning protects all cavity transitions. However, this conclusion holds only to lowest order in $\chi/\Delta_{bd}$. To estimate the range of protected cavity levels, we include higher-order corrections. We characterize the suppression of the susceptibility by the ratio
\begin{equation}
r_{n,m}\equiv \left| \frac{D_{n,m}^{\rm drive}}{D_{n,m}^{(0)}} \right|,
\end{equation}
where $D_{n,m}^{\rm drive}$ is the residual susceptibility at the sweet spot, $D_{n,m}^{(0)}$ is the corresponding susceptibility without the drive. We require $r_{n,m}$ to be smaller than the target tolerance $\epsilon$ for all $n,m<N$, where $N$ denotes the total number of the cavity levels that can be protected. This gives the estimate
\begin{equation}
N = \max \left( 1, \left\lfloor \frac{\epsilon}{3} \left| \frac{\Delta_{bd}}{\chi} \right| +\frac{1}{2} \right\rfloor \right),
\end{equation}
as detailed in App.~\ref{Perturbative Calculation of Cavity Transition frequencies}. Thus, the number of protected cavity levels is controlled by the small parameter $|\chi/\Delta_{bd}|$: a larger detuning relative to the dispersive shift protects more cavity levels. For the concrete example used in Sec.~\ref{sec4}, with $\Delta_{bd}/2\pi = 30~\mathrm{MHz}$, $\epsilon = 0.2$, and $\chi/2\pi = 0.5~\mathrm{MHz}$, we obtain $N = 4$.

\paragraph{Small-detuning regime.}
In this case, the FTT--drive detuning satisfies $|\Delta_{bd}|\ll |\chi|$.
In this limit, the shifted cavity transition frequency becomes
\begin{equation}
\widetilde{\omega}_{c;n,m}
\approx
(n-m)\bigg(\omega_c
- \frac{g^2}{\Delta_{bc}}\bigg)
+\delta_{m,0}\frac{\Omega_0^2}{\Delta_{bd}}.
\end{equation}
Differentiating with respect to $\Phi$ yields
\begin{equation}
\label{eq:Dm_selective}
D_{n,m}
\approx
\left[
(n-m)\frac{g^2}{\Delta_{bc}^2}
- \delta_{m,0}\frac{\Omega_0^2}{\Delta_{bd}^2}
\right]
\frac{\partial \omega_b}{\partial \Phi}.
\end{equation}
Therefore, a sweet spot can be generated only for a specific pair of Fock states. For example, for the transition with $n=1$ and $m=0$, choosing
\begin{equation}
\label{ss_selective}
\Delta_{bd}
= \pm
\left|
\frac{\Delta_{bc}}{g}\Omega_0
\right|
\end{equation}
generates a sweet spot. In practice, the detuning cannot be made arbitrarily small. Its lower bound is set by the finite linewidth of the FTT transition, characterized by the coherence time \(T_2\). To ensure that the drive remains outside the broadened FTT resonance, we require $|\Delta_{bd}|\gg 1/T_2$.

\paragraph{Sweet-spot contours in the space of drive parameters.}
We numerically evaluate the susceptibility $D_{1,0}$ as a function of the
drive amplitude $\Omega_0$ and the FTT--drive detuning $\Delta_{bd}$. The detuning is varied by sweeping the drive frequency.
The susceptibility is computed as the flux derivative of the
quasienergy difference, with the quasienergies obtained
using \texttt{QuTiP}~\cite{qutip5} (see
App.~\ref{app:numerics} for details). The results are shown in
Fig.~\ref{fig:compare}. The magnitude of the dispersive shift, \(|\chi|/2\pi \approx 2~\mathrm{MHz}\), sets the crossover between the small- and large-detuning regimes. This magnitude is comparable to values used in circuit-QED transmon--cavity experiments~\cite{Reed2010JCNReadoutdispersive,Premaratne2017FockSTIRAPdispersive,Heeres2017UniversalGateSetdispersive}. In the small-detuning regime, $|\Delta_{bd}|\ll|\chi|$, the sweet-spot condition in Eq.~\eqref{ss_selective} predicts the linear scaling $\Omega_0 \propto |\Delta_{bd}|$.
This scaling appears as two symmetric sweet-spot contours on either side of the effective resonance.
As the detuning increases, the left contour enters a regime with $|\Omega_0/\Delta_{bd}|>1$, and the perturbative approximation underlying Eq.~\eqref{energy correction} no longer applies.
By contrast, the right contour follows the superlinear scaling $\Omega_0 \propto \Delta_{bd}^{3/2}$ predicted by the large-detuning sweet-spot condition in Eq.~\eqref{eq:ss_unselective}. In a separate numerical example discussed in App.~\ref{snail}, we use SAFE to generate a sweet spot for a cavity coupled to a SNAIL.
We have also checked that realistic drive-amplitude fluctuations have
a negligible effect on cavity dephasing, as discussed in App.~\ref{app:amplitude_noise}.

For the remainder of this work, we focus on the large-detuning regime, which is more relevant to bosonic encodings because such encodings generally involve superpositions of multiple Fock states and therefore require simultaneous protection of several cavity transitions, as discussed later in Sec.~\ref{sec4}.

\section{Cavity Pure-Dephasing Channels}
\label{sec3}

While SAFE suppresses the dominant cavity dephasing induced by low-frequency flux fluctuations, additional cavity dephasing channels arise from other decoherence mechanisms. In this section, we evaluate the rates of these pure-dephasing channels. 

Our decoherence model treats two classes of noise differently. Relaxation and thermal excitation are described by Lindblad dissipators, whereas low-frequency ($1/f$) flux noise is treated explicitly as a stochastic contribution to the Hamiltonian. This distinction is necessary because the long temporal correlations of $1/f$ noise are not generally captured by a time-local Lindblad dissipator.

The dynamics are therefore modeled by a stochastic Lindblad master equation. For a given realization of the flux noise $\delta\Phi(t)$, the density matrix evolves as
\begin{align}
\label{Lindblad}
\dot{\rho} ={}& -i\left[H_{\mathrm{sys}}+H_{\text{drive}}(t)+H_{\mathrm{noise}}(t),\rho\right] \nonumber \\
&+ \gamma_{c\downarrow}\mathcal{D}[c]\rho
+ \gamma_{c\uparrow}\mathcal{D}[c^\dagger]\rho
+ \gamma_{b\downarrow}\mathcal{D}[b]\rho
+ \gamma_{b\uparrow}\mathcal{D}[b^\dagger]\rho ,
\end{align}
with the dissipator
\begin{equation}
\mathcal{D}[O]\rho = O\rho O^\dagger - \frac{1}{2}\left\{O^\dagger O,\rho\right\}.
\end{equation}
The rates $\gamma_{k\downarrow}$ and $\gamma_{k\uparrow}$, with $k\in\{b,c\}$, represent the total relaxation and thermal excitation rates of the respective subsystems.

\subsection{Effective Lindblad Master Equation}
\label{subsec:unselective-dephasing}
We obtain the effective drive Hamiltonian by a Schrieffer--Wolff transformation and derive the effective master equation shown in Eq.~\eqref{fullLME}; see App.~\ref{Transformation of the Dissipators and the Noise Hamiltonian}. To proceed, we truncate the FTT to a two-level system. This approximation is physically justified by three conditions.  First, when the FTT is used as an ancilla or coupler, it is initialized in its ground state during idle periods to minimize decoherence. Second, the applied drive is weak and far detuned from the FTT transition, $\Omega_0 \ll \Delta_{bd}$, so that the FTT has a negligible excited-state population. Finally, at typical dilution-refrigerator temperatures, $T \approx 10\text{--}20$~mK, the thermal energy is much smaller than the FTT frequency, $k_B T \ll \hbar \omega_b$. Consequently, the Boltzmann probability of populating higher excited states is vanishingly small.

After applying the truncation and secular approximations described in App.~\ref{Effective Lindblad Master Equation in a Truncated Subspace}, we obtain the effective master equation in the regime $\Delta_{bd}\gg\chi$:
\begin{align}
\label{transformed LME}
\dot{\rho}_{d} =& -i \left[ H_d + H_{\mathrm{noise},d}(t), \rho_d \right]
+ \widetilde{\gamma}_{c\downarrow} \mathcal{D}[c]\rho_d
+ \widetilde{\gamma}_{b\downarrow}\mathcal{D}[\sigma_-]\rho_d \nonumber \\
&+ \widetilde{\gamma}_{b\phi}\mathcal{D}[\sigma_+\sigma_-]\rho_d
+ \widetilde{\gamma}_{c\uparrow} \mathcal{D}[c^\dagger]\rho_d
+ \widetilde{\gamma}_{b\uparrow}  \mathcal{D}[\sigma_+]\rho_d .
\end{align}
The effective cavity relaxation and excitation rates are $\widetilde{\gamma}_{c\downarrow}  = \gamma_{c\downarrow} + (g/\Delta_{bc})^2\gamma_{b\downarrow}$ and $\widetilde{\gamma}_{c\uparrow}  = \gamma_{c\uparrow} + (g/\Delta_{bc})^2\gamma_{b\uparrow}$, respectively. Both rates include the inverse-Purcell effect. The FTT relaxation and excitation rates are $\widetilde{\gamma}_{b\downarrow}  = \gamma_{b\downarrow} [1-2(\Omega_0/\Delta_{bd})^2]$ and $\widetilde{\gamma}_{b\uparrow}  = \gamma_{b\uparrow} + \gamma_{b\downarrow}(\Omega_0/\Delta_{bd})^4$, respectively. The former rate is renormalized by the drive, whereas the latter contains thermal and drive-induced excitation contributions. Finally, $\widetilde{\gamma}_{b\phi}  = 4\gamma_{b\downarrow}(\Omega_0/\Delta_{bd})^2$ is the rate of drive-induced pure dephasing of the FTT.

The effective Hamiltonian and noise Hamiltonian in the new frame are
\begin{align}
H_d &\approx \left( \overline{\omega}_c + \chi \frac{\Omega_0^2}{\Delta_{bd}^2} \right) c^\dagger c + \left( \Delta_{bd} + \frac{2\Omega_0^2}{\Delta_{bd}} \right) \sigma_+\sigma_- \nonumber\\
&\quad + \chi \left( 1 - \frac{2\Omega_0^2}{\Delta_{bd}^2} \right) c^\dagger c \sigma_+\sigma_- , \\
H_{\mathrm{noise},d}(t) &\approx \delta\omega_b(t) \Bigg\{ -\frac{\Omega_0}{\Delta_{bd}}\sigma_x + \left[ 1 - 2\left(\frac{\Omega_0}{\Delta_{bd}}\right)^2 \right]\sigma_+\sigma_- \nonumber \\
&\quad + \left( \frac{g^2}{\Delta_{bc}^2} - 2\chi\frac{\Omega_0^2}{\Delta_{bd}^3} \right) c^\dagger c \nonumber \\
&\quad + \left( 4\chi\frac{\Omega_0^2}{\Delta_{bd}^3} - \frac{4Kg^2}{\Delta_{bc}^3} \right)\sigma_+\sigma_- c^\dagger c \Bigg\}.
\end{align}
The effective Hamiltonian $H_d$ contains the drive-induced corrections to the cavity frequency, the FTT transition frequency, and the dispersive shift. The noise Hamiltonian $H_{\mathrm{noise},d}$ describes the corresponding fluctuations of these corrected parameters induced by flux noise, together with a drive-induced transverse term that causes FTT depolarization.

\subsection{Total cavity pure-dephasing rate.}
There are three contributions to cavity pure dephasing: photon-shot noise, direct longitudinal dephasing of the cavity transition, and dephasing caused by fluctuations in the dispersive shift. We show below that this last contribution is negligible.

The first contribution, photon-shot noise, arises from FTT excitations, which dephase the cavity through the dispersive shift~\cite{shotnoise,PhysRevA.75.042302Ashphotonshot}. The FTT excitation rate has two contributions: thermal excitation at rate $\gamma_{b\uparrow}$, and drive-induced excitation from the transverse term ($\propto \sigma_x$) in the noise Hamiltonian. Under a Markovian approximation, the corresponding excitation rate is
\begin{equation}
\left(\frac{\Omega_0}{\Delta_{bd}}\right)^2 \left(\frac{\partial\omega_b}{\partial\Phi}\right)^2 S_{1/f}(\Delta_{bd}).
\end{equation}
In the strong-dispersive limit ($\chi\gg\gamma_{b\downarrow}$), the dephasing rate due to photon-shot noise is approximately equal to the FTT excitation rate~\cite{PhysRevA.75.042302Ashphotonshot}. Therefore, as shown below, photon-shot noise limits the cavity pure-dephasing rate. 

The second contribution is direct longitudinal dephasing, arising from the longitudinal component of $H_{\mathrm{noise},d}$, which induces fluctuations in the cavity frequency. This process leads to pure dephasing at the rate
\begin{equation}
A\left|\frac{\partial \omega_{b}}{\partial \Phi}
\left(
\frac{g^2}{\Delta_{bc}^2}
-2\chi\frac{\Omega_0^2}{\Delta_{bd}^3}
\right)\right|
\sqrt{2|\ln(\omega_\text{ir}t)|},
\end{equation}
where $\omega_\text{ir}$ is the infrared cutoff of the $1/f$ noise spectrum and $t$ is the experimental time. Following an earlier experiment that reported $\sqrt{|\ln(\omega_\text{ir}t)|}\approx 4$~\cite{PhysRevX.7.031037factor4}, we use this value in the calculations below. This longitudinal contribution can be suppressed at the dynamical sweet spot.

The third contribution arises from fluctuations in the dispersive shift when the FTT has excited-state population. In the regime considered here, however, the FTT is initialized in its ground state, and its excited-state population remains negligible. Using the effective relaxation and excitation rates introduced in Eq.~\eqref{transformed LME}, the corresponding effective population is given by
\begin{equation}
n_{\text{th}} = \frac{\widetilde{\gamma}_{b\uparrow}}{\widetilde{\gamma}_{b\downarrow}} = \frac{\gamma_{b\uparrow}}{\gamma_{b\downarrow}}+\left(\frac{\Omega_0}{\Delta_{bd}}\right)^4 + \mathcal{O}\left(\frac{\gamma_{b\uparrow}}{\gamma_{b\downarrow}}\left(\frac{\Omega_0}{\Delta_{bd}}\right)^2\right).
\end{equation}
Thus, $n_{\text{th}}$ is small. 

Consequently, the pure dephasing associated with fluctuations in the dispersive shift is negligible.
Collecting all the contributions, the cavity pure-dephasing rate is approximated by
\begin{align}
\label{eq:Gamma_phi_total_unselective}
\Gamma_{c\phi} &\approx A\left|\frac{\partial \omega_{b}}{\partial \Phi}\left( \frac{g^2}{\Delta_{bc}^2} - 2\chi\frac{\Omega_0^2}{\Delta_{bd}^3} \right)\right|\sqrt{2|\ln(\omega_\text{ir}t)|} \\
&\quad +\left(\frac{\Omega_0}{\Delta_{bd}}\right)^2 \left(\frac{\partial\omega_b}{\partial\Phi}\right)^2 \frac{A^2}{|\Delta_{bd}/2\pi|} + \frac{\Omega_0^4}{\Delta_{bd}^4} \gamma_{b\downarrow} + \gamma_{b\uparrow}. \nonumber
\end{align}
Substituting the dynamical sweet-spot condition from Eq.~\eqref{eq:ss_unselective} into Eq.~\eqref{eq:Gamma_phi_total_unselective}, we obtain the cavity pure-dephasing rate at the dynamical sweet spot:
\begin{equation}
\label{ssrate}
\Gamma_{c\phi}^{\mathrm{ss}}=\left(\frac{\partial\omega_b}{\partial\Phi}\right)^2\frac{A^2}{4|K/2\pi|} + \frac{\gamma_{b\downarrow}}{16}\left(\frac{\Delta_{bd}}{K}\right)^2 + \gamma_{b\uparrow}.
\end{equation}

\subsection{Suppression of Cavity Pure Dephasing}
\label{subsec:numerical_total_dephasing_rate}

\begin{figure*}[t]
\centering
\includegraphics[width=0.95\textwidth]{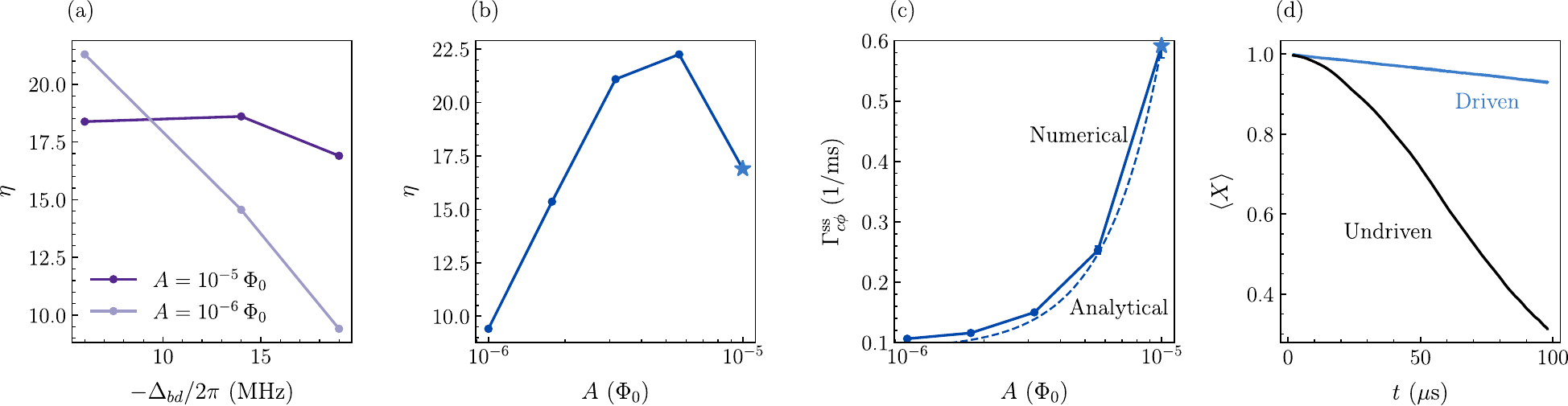}
\caption{Numerical evaluation of cavity pure-dephasing suppression and time-domain simulation of SAFE. \textbf{(a)} Pure-dephasing suppression factor $\eta$ as a function of the drive detuning $\Delta_{bd}$ for the $1/f$ flux-noise amplitudes $A=10^{-6}\Phi_0$ and $10^{-5}\Phi_0$.  SAFE suppresses the cavity pure-dephasing rate over the parameter range considered here. The suppression decreases more rapidly with $|\Delta_{bd}|$ at the smaller noise amplitude, whereas it remains nearly constant at moderate detunings for the larger noise amplitude. \textbf{(b)} $\eta$ as a function of the $1/f$ flux-noise amplitude $A$ at $\Delta_{bd}/2\pi=-19\ \text{MHz}$. The maximum suppression occurs near $A\approx 6\times 10^{-6}\Phi_0$. \textbf{(c)} Cavity pure-dephasing rates as functions of $A$ at $\Delta_{bd}/2\pi=-19\ \text{MHz}$. Solid lines show numerical results, and dashed lines show analytical estimates. 
\textbf{(d)} Time-domain decay of $\langle X(t)\rangle=\mathrm{Tr}\left[\bar{\rho}(t)X(t)\right]$ at $A=10^{-5}\Phi_0$ [marked by the star in (b) and (c)], where $X(t)=\ket{\phi_{g,1}(t)}\bra{\phi_{g,0}(t)}+\mathrm{h.c.}$. The extracted pure-dephasing times are $T_\phi\approx 100~\mu\mathrm{s}$ and $T_\phi\approx 2~\mathrm{ms}$ in the undriven and driven cases, respectively.}
\label{fig:total_rate}
\end{figure*}

To quantify the benefit of the dynamical sweet spot, we define the
pure-dephasing suppression factor
\begin{equation}
\eta
=
\frac{\Gamma_{c\phi}^{(0)}}{\Gamma_{c\phi}^{\mathrm{ss}}},
\end{equation}
where 
\begin{equation}
    \Gamma_{c\phi}^{(0)} = A\bigg|\frac{\partial \omega_{b}}{\partial \Phi}\bigg| \frac{g^2}{\Delta_{bc}^2}\sqrt{2|\ln(\omega_\text{ir}t)|} + \gamma_{b\uparrow}
\end{equation} 
is the cavity pure-dephasing rate without the
drive, and $\Gamma_{c\phi}^{\mathrm{ss}}$ is the residual pure-dephasing rate at the
dynamical sweet spot. Values of $\eta>1$ indicates suppression of the cavity pure-dephasing rate.

To suppress pure dephasing, we use a simple rule for choosing
the drive detuning. As discussed in Sec.~\ref{large-detuning regime}, the
desired number of protected cavity states sets a lower bound on
$|\Delta_{bd}|$: a larger detuning protects more number of cavity states. However, the residual rate $\Gamma_{c\phi}^{\mathrm{ss}}$ scales as $\gamma_{b\downarrow}(\Delta_{bd}/K)^2$, resulting in a trade-off between the number of protected states and the achievable suppression factor. As shown in Fig.~\ref{fig:total_rate}(a), $\eta$ generally decreases with increasing $|\Delta_{bd}|$. This decrease is particularly pronounced for $A=10^{-6}\Phi_0$, whereas for $A=10^{-5}\Phi_0$, $\eta$ remains nearly constant at moderate detunings before decreasing at larger $|\Delta_{bd}|$. This difference arises because the detuning-independent $A^2$ term dominates $\Gamma_{c\phi}^{\mathrm{ss}}$ at large $A$, whereas the term $\gamma_{b\downarrow}(\Delta_{bd}/K)^2$ becomes important earlier at small $A$. We therefore choose $|\Delta_{bd}|$ as small as possible while still protecting the desired number of cavity states.

Most device parameters affect $\eta$ in a relatively straightforward way. For example, within the present approximation, increasing $|K|$ or decreasing $\gamma_{b\downarrow}$ reduces $\Gamma_{c\phi}^{\mathrm{ss}}$ and therefore increases $\eta$. The dependence on the flux-noise amplitude $A$, however, can be more subtle. To illustrate this, we compute $\eta$ as a function of $A$ for a set of representative FTT--cavity parameters in the dispersive regime at zero temperature, see App.~\ref{app:numerical_dephasing_rates} for more details. The results are shown in Fig.~\ref{fig:total_rate}(b). Over this range, SAFE suppresses the cavity pure-dephasing rate by approximately one order of magnitude. For the parameters used here, the suppression factor reaches a maximum of $\eta\approx 22$ near $A\approx 6\times 10^{-6}\Phi_0$. This maximum appears because $\Gamma_{c\phi}^{\mathrm{ss}}$ has different scaling behaviors for small and large $A$, as shown in Fig.~\ref{fig:total_rate}(c). At small $A$, $\Gamma_{c\phi}^{\mathrm{ss}}$ in Eq.~\eqref{ssrate} is dominated by the $A$-independent contribution, while $\Gamma_{c\phi}^{(0)}$ scales linearly with $A$. This leads to $\eta \propto A$. At larger $A$, $\Gamma_{c\phi}^{\mathrm{ss}}$ grows approximately as $A^2$, while $\Gamma_{c\phi}^{(0)}$ scales linearly with $A$. This leads to $\eta \propto 1/A$.

We present a representative Ramsey simulation for $A=10^{-5}\Phi_0$ to visualize the suppression of cavity pure dephasing. The chosen initial state is
$\ket{\psi_0}=\big(\ket{g,0}+\ket{g,1}\big)/{\sqrt{2}}$.
When the drive is turned on adiabatically, the states $\ket{g,0}$ and $\ket{g,1}$ are mapped onto the target Floquet states $\ket{\phi_{g,0}(t)}$ and $\ket{\phi_{g,1}(t)}$, respectively; see App.~\ref{Adiabatic transition} for further discussion. The coherence time $T_2$ is extracted from the decay of 
\begin{align}
\langle X(t)\rangle &= \mathrm{Tr}\!\left[\bar{\rho}(t)\,X(t)\right],\\
X(t) &= \ket{\phi_{g,1}(t)}\!\bra{\phi_{g,0}(t)} + \mathrm{h.c.},
\end{align}
where $\bar{\rho}(t)$ is the density matrix expressed in a frame rotating at the cavity frequency and averaged over flux-noise trajectories. The coherence time $T_2$ contains contributions from both relaxation and pure dephasing. From a separate numerical simulation, we obtain a cavity relaxation time of $2.1~\mathrm{ms}$ both with and without the drive. Although a drive can modify the effective relaxation rates between Floquet states in general, this modification is a negligible higher-order effect in the SAFE scheme. The results for $\langle X(t)\rangle$ as a function of time are shown in Fig.~\ref{fig:total_rate}(d). Without the drive, the coherence exhibits the Gaussian decay profile expected for dephasing dominated by low-frequency $1/f$ flux noise. We extract $T_2$ as the time at which the coherence envelope decays to $1/e$ of its initial value and obtain $T_2\approx T_\phi\approx 0.1~\mathrm{ms}$. This indicates that the coherence is limited by pure dephasing. In the presence of the sweet-spot drive, the leading sensitivity to low-frequency flux noise is suppressed, and the residual decoherence channels are approximately Markovian, as discussed in Sec.~\ref{subsec:unselective-dephasing}. Over the simulated short-time window, the coherence exhibits approximately linear behavior, as expected from the short-time expansion $e^{-t/T_2}\approx 1-t/T_2$. We therefore fit the coherence decay to an exponential envelope and obtain $T_2\approx 1.4~\mathrm{ms}$, with an extracted pure-dephasing time of $T_\phi\approx 2~\mathrm{ms}$. Thus, the SAFE drive increases the pure-dephasing time from approximately $0.1~\mathrm{ms}$ to $2~\mathrm{ms}$, corresponding to a 20-fold improvement.

\section{Application to Bosonic Qubits}
\label{sec4}

\begin{figure*}[t]
\centering
\includegraphics[width=\textwidth]{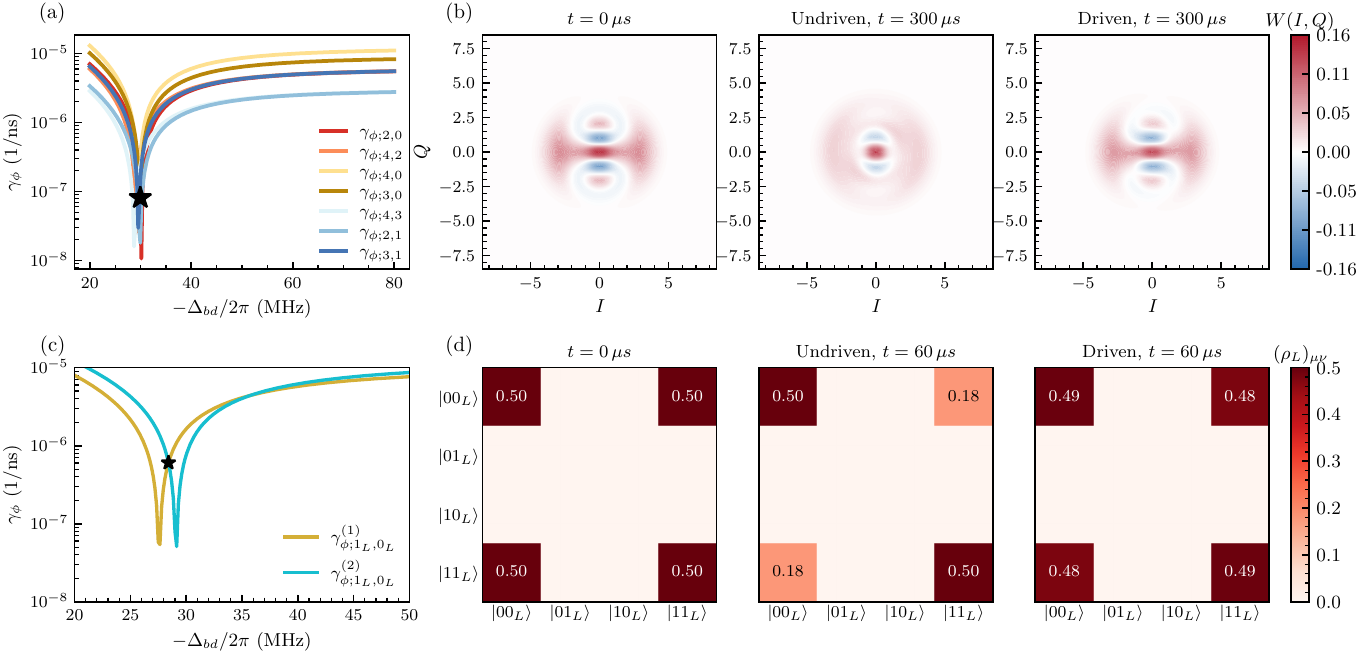}
\caption{Improvement of the coherence of bosonic encodings using SAFE
(the drive amplitude is $\Omega_0/2\pi=10~\mathrm{MHz}$ throughout).
\emph{Binomial code}---\textbf{(a)}~Dephasing rates ($\gamma_{\phi;m,n}$) induced by FTT frequency fluctuations as a function of
$\Delta_{bd}$. The star marks $\Delta_{bd}/2\pi\approx-30~\mathrm{MHz}$, where all
relevant $\gamma_{\phi;m,n}$ are suppressed by factors of at least 13.
\textbf{(b)}~Wigner function of $\ket{+_L}$ at $t=300~\mu\mathrm{s}$, plotted in the dimensionless phase-space coordinates $I$ and $Q$. The driven case retains the interference features
that are lost without the drive. The decoherence-induced infidelity is reduced from
$0.3$ without the drive to $0.043$ with the drive.
\emph{Dual-rail encoding}---\textbf{(c)}~Dephasing rate of
each dual-rail qubit as a function of
$\Delta_{bd}$. Both qubits exhibit approximately tenfold lower dephasing rates near $\Delta_{bd}/2\pi\approx -29~\mathrm{MHz}$.
\textbf{(d)}~Logical density-matrix elements for two dual-rail qubits initialized in a
logical Bell state. SAFE suppresses dephasing, while leakage from single-photon loss
remains unchanged. The decoherence-induced infidelity is reduced from $0.18$ without
the drive to $0.011$ with the drive. The system parameters used in the simulations are
given in App.~\ref{System Parameters Used in the simulations}.}
\label{fig:qec_plot}
\end{figure*}

Logical states can be encoded in the cavity modes of a cavity--FTT architecture, in which the FTT serves as an ancilla or coupler for control and recovery operations. In such systems, however, the cavity inherits dephasing from the FTT. We show that SAFE can suppress pure dephasing across multiple Fock-state pairs and multiple cavities, using binomial and dual-rail qubits as representative examples.

\subsection{Binomial Qubit}
We consider the minimal binomial encoding~\cite{PhysRevX.6.031006Binonmial}
\begin{equation}
\ket{0_L}=\frac{\ket{0}+\ket{4}}{\sqrt{2}},\qquad \ket{1_L}=\ket{2},
\end{equation}
which is designed to correct a single-photon-loss error in a QEC protocol.

In this encoding, pure dephasing of several Fock-state coherences plays an important role. Within the code space, pure dephasing of the $\ket{0}$--$\ket{4}$ coherence distorts $\ket{0_L}$.
Moreover, pure dephasing of the coherences associated with the $\ket{0}$--$\ket{2}$, $\ket{0}$--$\ket{4}$, and $\ket{2}$--$\ket{4}$ pairs contributes to logical errors. Single-photon loss leads to the transitions $\ket{0_L} \rightarrow \ket{3}$ and $\ket{1_L} \rightarrow \ket{1}$. The logical information temporarily resides in the error subspace $\{\ket{0_E}=\ket{3}, \ket{1_E}=\ket{1}\}$ before a recovery operation is applied. Preserving the $\ket{1}$--$\ket{3}$ coherence is therefore important. During recovery from the error subspace, the system can occupy superpositions of logical and error states. This makes pure dephasing of the $\ket{0}$--$\ket{3}$, $\ket{3}$--$\ket{4}$, and $\ket{1}$--$\ket{2}$ coherences crucial.

The logical states are encoded in the cavity mode of a cavity--FTT system, with the FTT serving as an ancilla for control.
We apply SAFE to suppress the cavity pure dephasing inherited from the FTT. To identify suitable SAFE drive parameters, we sweep the detuning $\Delta_{bd}$ at a fixed drive amplitude $\Omega_0$ and evaluate the pure-dephasing rate using
\begin{equation}
\gamma_{\phi;m,n}
= A\bigg|\frac{\partial \widetilde{\omega}_{c;m,n}}{\partial \Phi}\bigg|
\sqrt{2|\ln(\omega_\text{ir}t)|}.
\end{equation}
Here, $\widetilde{\omega}_{c;m,n}$ is the transition frequency between Fock states $m$ and $n$ when the FTT is in its ground state. Figure~\ref{fig:qec_plot}(a) shows the rates
$\gamma_{\phi;m,n}$ for several Fock-state coherences relevant to the binomial code.
Each dephasing rate exhibits a minimum as a function of the drive detuning.
The minima occur near the same detuning, with the chosen operating point marked by the black star at $\Delta_{bd}/2\pi \approx -30\,\mathrm{MHz}$. At this point, SAFE suppresses the pure dephasing of several coherences in the code and error subspaces by about an order of magnitude. This is expected because, in the large-detuning regime, the ratio $|\Delta_{bd}/\chi|\approx 60\gg 1$ reduces the spread among the minima, as discussed in Sec.~\ref{sec2}.

To visualize how SAFE protects logical coherence against pure dephasing in phase space, we numerically solve the Lindblad master equation in Eq.~\eqref{Lindblad} with the initial state $\ket{\overline{+}_L} = (\ket{\overline{0}_L}+\ket{\overline{1}_L})/\sqrt{2}$. Here we use dispersively dressed states to define logical states
\[
\ket{\overline{0}_L}=\frac{\ket{\overline{g,0}}+\ket{\overline{g,4}}}{\sqrt{2}},
\qquad
\ket{\overline{1}_L}=\ket{\overline{g,2}},
\]
. We then compute the Wigner functions of the initial state and the final states at $t=300~\mu\mathrm{s}$ as shown in Fig.~\ref{fig:qec_plot}(b).
Without the drive, the Wigner function after $300\,\mu\mathrm{s}$ is strongly distorted relative to the initial state, and the interference features associated with the logical superposition are substantially reduced. By contrast, with the SAFE drive, the overall phase-space structure is better preserved and the interference features remain largely visible, providing a visual indication that SAFE protects coherence in the logical subspace.

We quantify the decoherence-induced error using the state-transfer infidelity
\begin{equation}
I(t)=1-\sqrt{\bra{\psi_{\mathrm{id}}(t)}\bar{\rho}(t)\ket{\psi_{\mathrm{id}}(t)}} .
\end{equation}
Here, $\bar{\rho}(t)$ is the density matrix averaged over flux-noise trajectories, whereas $\ket{\psi_{\mathrm{id}}(t)}$ is obtained by evolving the same initial state under the closed-system Hamiltonian.
By using $\ket{\psi_{\mathrm{id}}(t)}$ as the reference state, the infidelity does not count the coherent evolution generated by the Hamiltonian as an error, but instead quantifies the deviation caused by decoherence. After $300\,\mu\mathrm{s}$, SAFE improves the infidelity from $0.3$ to $0.043$ by suppressing pure dephasing.

\subsection{Dual-Rail Qubit}
\begin{figure}[t]
    \centering
    \includegraphics[width=1\linewidth]{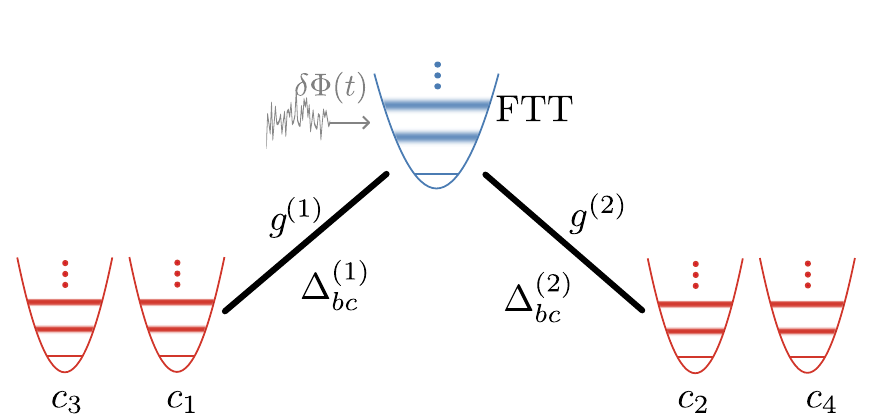}
    \caption{Schematic of the system of two dual-rail qubits considered in this work. The first logical qubit is encoded in cavities $c_1$ and $c_3$, and the second logical qubit is encoded in cavities $c_2$ and $c_4$. The rails $c_1$ and $c_2$, one from each dual-rail qubit, are dispersively coupled to a common flux-tunable transmon (FTT) ancilla with respective coupling strengths $g^{(1)}$ and $g^{(2)}$ and detunings $\Delta_{bc}^{(1)}$ and $\Delta_{bc}^{(2)}$. Flux fluctuations $\delta\Phi(t)$ modulate the FTT frequency and induce correlated frequency fluctuations in the coupled cavity rails through the dispersive Lamb shift.}
    \label{fig:dualrail}
\end{figure}

We next consider a second bosonic encoding in which photon loss is converted into a detectable erasure error: the dual-rail code~\cite{dualrailcavity1,teoh2023dual}.
A single dual-rail qubit can be encoded in two cavities with logical states $\ket{0_L}=\ket{01}$ and $\ket{1_L}=\ket{10}$. 
In the following, we consider two dual-rail qubits, as shown in Fig.~\ref{fig:dualrail}. In the simplified model used below, one rail of each dual-rail qubit is dispersively coupled to the same FTT ancilla, while the remaining rails are not directly coupled to the FTT. We assume that the FTT-coupled rails are mutually far detuned to suppress FTT-mediated cavity--cavity interaction, while an appropriate choice of cavity frequencies renders the residual cross-Kerr interactions negligible~\cite{Zhu2013}. Additional couplers required for operations on a single dual-rail qubit are not modeled explicitly.

In architectures where the FTT is biased away from its flux sweet spot, e.g., to enable parametric three-wave-mixing interactions, flux noise induces fluctuations of the FTT frequency. As shown in Sec.~\ref{sec2}, these fluctuations are inherited by the cavities through the dispersive Lamb shift, giving
\begin{equation}
\delta \widetilde{\omega}^{(l)}_{c}(t)=
\biggl(\frac{g^{(l)}}{\Delta^{(l)}_{bc}}\biggr)^2
\delta\omega_b(t).
\end{equation}
Here, $g^{(l)}$ and $\Delta^{(l)}_{bc}$ are, respectively, the coupling strength and detuning between the FTT and cavity $l$, for $l=1,2$. These fluctuations contribute directly to logical dephasing of the corresponding dual-rail qubits.

We choose the SAFE drive frequency and amplitude to reduce the flux-noise sensitivity of both FTT-coupled cavity modes.
As shown in Fig.~\ref{fig:qec_plot}(c), the two minima occur in the same detuning window, and there is a common detuning near $-\Delta_{bd}/2\pi\approx 29\,\mathrm{MHz}$ at which both dephasing rates are reduced by an order of magnitude. A naive application of Eq.~\eqref{eq:ss_unselective} would suggest that the sweet-spot location is independent of the cavity parameters and should therefore be identical for both cavities. Higher-order corrections, however, lift this degeneracy, as shown in App.~\ref{Multimode Sweet-Spot Alignment}. The spread between the two minima scales as $\Omega_0^{4/3}$, indicating that optimizing the drive amplitude can further suppress pure dephasing due to flux noise.

To illustrate the effect of this dephasing suppression on an encoded logical state, we numerically evolve the density matrix under the Lindblad master equation with the initial Bell state
\begin{equation}
\ket{\psi_{\mathrm{Bell}}}=\frac{1}{\sqrt{2}}\left(\ket{00_L}+\ket{11_L}\right).
\end{equation}
Figure~\ref{fig:qec_plot}(d) shows the logical density matrix in frames rotating at the respective dressed frequencies of the cavity modes. We plot $(\rho_L)_{\mu\nu}$ for $\mu,\nu\in\{00_L,01_L,10_L,11_L\}$ within the logical subspace.
Without SAFE, the off-diagonal matrix elements $(\rho_L)_{00_L,11_L}$ and $(\rho_L)_{11_L,00_L}$ decay rapidly. With SAFE, these coherence magnitudes remain substantially larger over the same time interval, showing that SAFE suppresses logical dephasing. The reduced values of the logical-subspace diagonal elements reflect population leakage out of the logical subspace into the physical ground state of a dual-rail pair. The photon-loss rate remains essentially unchanged under the SAFE drive, indicating that SAFE has a negligible effect on the erasure error and therefore can help preserve the photon-loss bias. Quantitatively, the decoherence-induced infidelity improves from $0.18$ without SAFE to $0.011$ with SAFE.

\section{Conclusion}
\label{sec5}

In summary, we have introduced the SAFE technique, which dynamically suppresses cavity dephasing inherited from a coupled nonlinear mode subject to \(1/f\) noise. Our numerical results show that SAFE improves the cavity dephasing time by more than an order of magnitude in a representative example. Meanwhile, SAFE has a negligible effect on the cavity relaxation time. 
We further illustrate the potential of SAFE for binomial and dual-rail qubits, highlighting its ability to suppress dephasing simultaneously across multiple pairs of Fock states and multiple cavity modes.
Overall, SAFE provides a practical route toward the long coherence times needed for bosonic encodings and quantum memories.


\section*{Acknowledgments}
We thank Rohan Narayan Rajmohan, Sai Paivan Chitta, Tianpu Zhao, Ziqian Li, and Eesh Gupta for illuminating discussions. We thank David I. Schuster for supporting the experimental exploration that helped shape this work. This material is based upon work supported by the U.S. Department of Energy, Office of Science, National Quantum Information Science Research Centers, through the Co-design Center for Quantum Advantage (C2QA) under Contract No. DE-SC0012704, and the Superconducting Quantum Materials and Systems Center (SQMS) under Contract No. DE-AC02-
07CH11359. Yao Lu acknowledges support from the DOE
Early Career Research Program.

\section*{Data Availability}
The data supporting the findings of this article were generated by numerical
simulations. The source code, parameters, and data used to generate the figures
are publicly available in the archived SAFE release~\cite{SAFEDataRelease2026}.
\appendix
\section{Floquet Formalism}
\label{app:floquet}
In this appendix, we justify using the static rotating-frame Hamiltonian $H_R$ [Eq.~\eqref{eq:H_R_floquet}] to approximate the quasienergies and Floquet states of the periodically driven dispersive Hamiltonian.
\subsection{Floquet Theorem}
Consider a Hilbert space $\mathcal{H}$ of dimension $N$ and a time-dependent Hamiltonian $H(t)$ with period $\tau$, satisfying $H(t+\tau) = H(t)$. Floquet's theorem states that the solutions to the SchrÃ¶dinger equation can be written as
\begin{equation}
\ket{\phi_n(t)} = e^{-i\varepsilon_n t} \ket{\psi_n(t)},
\end{equation}
where $\{\ket{\psi_n(t)}\}$ is a $\tau$-periodic orthonormal basis and $\{\varepsilon_n\}$ is the set of quasienergies.
\subsection{Construction of the Solution}
\label{Construction of the Solution}
To obtain a differential equation relating the periodic states $\ket{\psi_n(t)}$ and the quasienergies $\varepsilon_n$ to the Hamiltonian $H(t)$, we substitute the ansatz $\ket{\Psi(t)} = e^{-i\varepsilon_n t} \ket{\psi_n(t)}$ into the time-dependent SchrÃ¶dinger equation. This yields the Floquet eigenvalue equation
\begin{equation}
\label{quasienergy equation}
\bigl(H(t)-i\partial_{t}\bigr) \ket{\psi_n(t)} = \varepsilon_n \ket{\psi_n(t)}.
\end{equation}
Since the quantities on both sides of Eq.~\eqref{quasienergy equation} are periodic, we can expand them in Fourier series. A concise way to formalize this expansion is to introduce the \emph{extended Hilbert space} $\mathcal{F}$ as
\begin{equation}
\mathcal{F} = \mathrm{span}\bigl\{ \kett{\mu,n} \big| \mu\in\mathbb{Z}, n=1,\dots,N \bigr\}.
\end{equation}
We use the double-ket notation $\kett{\cdot}$ to denote a basis vector in this extended space. Each basis vector corresponds to a function that maps time $t \in \mathbb{R}$ to a state in the physical Hilbert space $\mathcal{H}$:
\begin{equation}
\kett{\mu,n} : \mathbb{R} \rightarrow \mathcal{H}.
\end{equation}
Specifically, these basis states are defined as
\begin{equation}
\kett{\mu,n}(t) = \ket{n} e^{i\mu\omega_d t}, \qquad \omega_d \equiv \frac{2\pi}{\tau},
\end{equation}
where $\{\ket{n}\}$ is an orthonormal basis of $\mathcal{H}$. The inner product on $\mathcal{F}$ is defined by averaging over one period:
\begin{align}
\bbrakett{\mu',n'}{\mu,n}
&\equiv \frac{1}{\tau}\int_{0}^{\tau}
\bigl[\kett{\mu',n'}(t)\bigr]^\dagger \kett{\mu,n}(t) dt \nonumber \\
&= \frac{1}{\tau}\int_{0}^{\tau} e^{-i\mu'\omega_d t}\bra{n'}\ket{n}e^{i\mu\omega_d t} dt \nonumber \\
&= \delta_{n,n'} \frac{1}{\tau}\int_{0}^{\tau} e^{i(\mu-\mu')\omega_d t} dt \nonumber \\
&= \delta_{\mu,\mu'} \delta_{n,n'}.
\end{align}
To rewrite Eq.~\eqref{quasienergy equation} in the basis $\{\kett{\mu,n}\}$, we define the quasienergy operator $\overline{Q}:\mathcal{F}\to\mathcal{F}$, whose action in the time domain is
\begin{equation}
\bigl(\overline{Q} \kett{f}\bigr)(t) = \bigl(H(t)-i\partial_{t}\bigr) \ket{f(t)}.
\end{equation}
We define the vector $\kett{\psi_n} \in \mathcal{F}$ corresponding to the periodic function $\ket{\psi_n(t)}$ as
\begin{equation}
\label{eigenvector definition}
\kett{\psi_n}(t) = \ket{\psi_n(t)}.
\end{equation}
Consequently, Eq.~\eqref{quasienergy equation} can be represented as an eigenvalue problem in the extended Hilbert space:
\begin{equation}
\label{evals equation}
\overline{Q} \kett{\psi_n} = \varepsilon_n \kett{\psi_n}.
\end{equation}
To compute the eigenvalues and eigenvectors, we construct the matrix representation of $\overline{Q}$ in the basis $\{\kett{\mu,n}\}$:
\begin{equation}
\label{Qbar}
\overline{Q} = \sum_{\mu',n'}\sum_{\mu,n} \bar{q}_{\mu' n',\mu n} \kett{\mu',n'}\bbra{\mu,n},
\end{equation}
with matrix elements
\begin{align}
\label{smallq}
\bar{q}_{\mu' n',\mu n}
&= \bbrakett{\mu',n'}{\overline{Q}|\mu,n} \nonumber \\
&= \frac{1}{\tau}\int_{0}^{\tau} \bigl[\kett{\mu',n'}(t)\bigr]^\dagger \bigl(\overline{Q} \kett{\mu,n}(t)\bigr) dt.
\end{align}
Substituting the explicit form $\kett{\mu,n}(t)=\ket{n}e^{i\mu\omega_d t}$ gives
\begin{align}
\bar{q}_{\mu' n',\mu n}
&= \frac{1}{\tau}\int_{0}^{\tau} e^{-i\mu'\omega_d t} \bra{n'} \bigl(H(t)-i\partial_t\bigr) \bigl(\ket{n}e^{i\mu\omega_d t}\bigr) dt \nonumber \\
&= \frac{1}{\tau}\int_{0}^{\tau} \langle n'|H(t)|n\rangle e^{i(\mu-\mu')\omega_d t} dt + \mu \omega_d \delta_{\mu'\mu} \delta_{n'n} \nonumber \\
&= h_{n' n}^{(\mu-\mu')} + \mu \omega_d \delta_{\mu'\mu} \delta_{n'n},
\end{align}
where $h_{n'n}^{(k)}$ are the Fourier components of the Hamiltonian:
\begin{equation}
H^{(k)} = \frac{1}{\tau}\int_{0}^{\tau} H(t) e^{ik\omega_d t} dt, \qquad h^{(k)}_{n' n}=\langle n'|H^{(k)}|n\rangle.
\end{equation}
Thus, the operator $\overline{Q}$ can be written as a block matrix, with the $\mu$ sectors ordered, for example, from $-2$ to $2$:
\begin{widetext}
\begin{equation}
\bar{q} = \begin{pmatrix}
\ddots & & & & & \\
& h^{(0)}-2\omega_d I & h^{(1)} & h^{(2)} & h^{(3)} & h^{(4)} \\
& h^{(-1)} & h^{(0)}-\omega_d I & h^{(1)} & h^{(2)} & h^{(3)} \\
& h^{(-2)} & h^{(-1)} & h^{(0)} & h^{(1)} & h^{(2)} \\
& h^{(-3)} & h^{(-2)} & h^{(-1)} & h^{(0)}+\omega_d I & h^{(1)} \\
& h^{(-4)} & h^{(-3)} & h^{(-2)} & h^{(-1)} & h^{(0)}+2\omega_d I \\
& & & & & \ddots
\end{pmatrix},
\end{equation}
\end{widetext}
where $I$ is the identity matrix. Finally, the quasienergies are found by solving the characteristic equation
\begin{equation}
\det\bigl(\bar{q}-\varepsilon \bar{I}\bigr)=0,
\end{equation}
where $\bar{I}$ is the identity operator in the extended Hilbert space.
\subsection{Quasienergies and Floquet States of the Driven Dispersive Hamiltonian}
In this section, we use the procedure described above to approximate the quasienergies and Floquet states of the driven dispersive Hamiltonian given by
\begin{equation}
\begin{split}
\label{Driven dispersive}
H(t) &= \bar{\omega}_b b^\dagger b + \frac{K}{2} b^{\dagger 2} b^2 + \bar{\omega}_c c^\dagger c \\
&\quad + \chi b^\dagger b c^\dagger c + 2\Omega_0 \cos(\omega_d t) (b + b^\dagger).
\end{split}
\end{equation}
\subsubsection{Quasienergy Operator for the Driven Dispersive Hamiltonian}
To obtain the quasienergies, we first construct $\overline{Q}$ in the extended Hilbert space using the basis states $\kett{\mu,(n,m)}$, where $\kett{\mu,(n,m)}(t) = e^{i\mu\omega_d t}\ket{n,m}$ maps each time to a basis state of the static dispersive Hamiltonian. Using Eq.~\eqref{smallq}, we obtain
\begin{equation}
\overline{Q} = \overline{Q}_0 + \overline{V},
\end{equation}
with
\begin{widetext}
\begin{align}
\overline{Q}_0 &= \sum_{\mu,n,m} \left( \bar{\omega}_b n + \frac{K}{2} n(n-1) + \bar{\omega}_c m + \chi nm + \mu \omega_d \right)
\kett{\mu,(n,m)}\bbra{\mu,(n,m)}, \\
\overline{V} &= \Omega_0 \sum_{\mu,n,m} \sqrt{n+1}
\left( \kett{\mu+1,(n+1,m)}\bbra{\mu,(n,m)}
+ \kett{\mu-1,(n+1,m)}\bbra{\mu,(n,m)} \right)
+ \text{h.c.}
\end{align}
\end{widetext}
Here, the unperturbed component $\overline{Q}_0$ is diagonal in the basis $\{\kett{\mu,(n,m)}\}$, and $\overline{V}$ arises from the periodic drive.
In the extended Hilbert space, the drive connects adjacent sectors $\mu\to\mu\mp 1$ and excites the FTT from $n\to n\pm1$. These processes have detunings
\begin{align}
\Delta E_1 &= \bar{\omega}_b - \omega_d + (n-1)K + \chi m,\\
\Delta E_2 &= \bar{\omega}_b + \omega_d+ (n-1)K + \chi m,
\end{align}
respectively. In the parameter regime of interest, $|\bar{\omega}_b - \omega_d|\ll\omega_d$, we have the hierarchy
\begin{equation}
\label{hierarchy}
\Omega_0 < |\Delta E_1| \ll |\Delta E_2|,
\end{equation}
so that $\overline{V}$ can be treated perturbatively.
\subsubsection{Approximation of Quasienergies and Floquet States}
To show that the eigenvalues of $H_R$ approximate the quasienergies, we block-diagonalize the operator $\overline{Q}$ via a Schrieffer--Wolff transformation (SWT) within the near-resonant subspace selected by the hierarchy in Eq.~\eqref{hierarchy}. Specifically, we define the projectors
\begin{equation}
\bar P_\mu=\sum_{n,m}\kett{\mu-n,(n,m)}\bbra{\mu-n,(n,m)},
\end{equation}
which collect the states most strongly coupled by the drive. Since $\bar P_\mu$ mixes different $\mu$ sectors, it is convenient to relabel the basis using the unitary
\begin{equation}
\bar R=\sum_{\mu,n,m}\kett{\mu,(n,m)}\bbra{\mu-n,(n,m)}.
\end{equation}
In this relabeled frame,
\begin{align}
\bar P_\mu^{R} &= \bar R^\dagger \bar P_\mu \bar R
=\sum_{n,m}\kett{\mu,(n,m)}\bbra{\mu,(n,m)},\\
\bar Q_{R} &= \bar R^\dagger \bar Q \bar R,
\end{align}
so that $\bar Q_R$ has the matrix representation
\begin{widetext}
\begin{equation}
\bar q_R=
\begin{pmatrix}
\ddots & \vdots & \vdots & \vdots & \vdots & \vdots & \ddots \\
\cdots & [H_R-2\omega_d I] & 0 & \Omega_0[b] & 0 & 0 & \cdots \\
\cdots & 0 & [H_R-\omega_d I] & 0 & \Omega_0[b] & 0 & \cdots \\
\cdots & \Omega_0[b^\dagger] & 0 & [H_R] & 0 & \Omega_0[b] & \cdots \\
\cdots & 0 & \Omega_0[b^\dagger] & 0 & [H_R+\omega_d I] & 0 & \cdots \\
\cdots & 0 & 0 & \Omega_0[b^\dagger] & 0 & [H_R+2\omega_d I] & \cdots \\
\ddots & \vdots & \vdots & \vdots & \vdots & \vdots & \ddots
\end{pmatrix}.
\end{equation}
\end{widetext}
where $[O]$ denotes the matrix of $O$ in the $\{\ket{n,m}\}$ basis. The leading-order approximation to the effective quasienergy operator is given by the block-diagonal part,
\begin{equation}
\sum_\mu \bar P_\mu^{R} \bar Q_R \bar P_\mu^{R}.
\end{equation}
For the drive parameters of interest, it suffices to diagonalize a single sector, e.g., $\mu=0$:
\begin{equation}
H_R\ket{\psi_j}=E_j\ket{\psi_j},
\qquad
\ket{\psi_j}=\sum_{n,m}c_{nm}^j\ket{n,m}.
\end{equation}
The corresponding quasienergy is then
\begin{equation}
\varepsilon_{j,\mu}=E_j+\mu\omega_d,
\end{equation}
and the corresponding Floquet state in the relabeled frame is
\begin{equation}
\kett{\psi_{j,\mu}}_R=\sum_{n,m}c_{nm}^j\kett{\mu,(n,m)}.
\end{equation}
\subsubsection{Connection with the RWA}
The relabeling $\bar R$ in the extended Hilbert space corresponds to moving to a frame that rotates with the drive in the physical Hilbert space. Applying $\bar R$ to a Floquet basis state gives
\begin{align}
\bigl(\bar R\kett{\mu,(n,m)}\bigr)(t)
&= \kett{\mu+n,(n,m)}(t)
= \ket{n,m} e^{i(\mu+n)\omega_d t} \nonumber\\
&= \bigl(e^{in\omega_d t}\ket{n,m}\bigr) e^{i\mu\omega_d t}.
\end{align}
Thus, $\bar R$ amounts to multiplying $\ket{n,m}$ by the phase $e^{in\omega_d t}$. This is precisely the action of the rotating-frame unitary
\begin{equation}
U_R(t)=e^{-i\omega_d t b^\dagger b},
\end{equation}
which yields the effective Hamiltonian
\begin{align}
H_{\mathrm{eff}}(t)&=U_R^\dagger(t)H(t)U_R(t)-i U_R^\dagger(t)\dot U_R(t)\nonumber\\
&= H_R +\Omega_0(e^{-i2\omega_dt}b + \text{h.c.}).
\end{align}
In this frame, the terms oscillating at $\pm 2\omega_d$ vary rapidly. Applying the RWA gives the time-independent Hamiltonian $H_R$. This approximation corresponds to the first-order SWT discussed above.
\section{Perturbative Calculation of Cavity Transition Frequencies}
\label{Perturbative Calculation of Cavity Transition frequencies}
In this section, we first perform a systematic fourth-order perturbative calculation to confirm the sweet-spot condition in the large-detuning regime discussed in the main text. We then carry out higher-order perturbative calculations to estimate how many levels can be protected by the SAFE scheme.
We start from the rotating-frame Hamiltonian including fourth-order corrections,
\begin{equation}
\begin{aligned}
H_R &= \Delta_{bd} b^\dagger b + \frac{{K}}{2} b^{\dagger 2} b^2 + \bar{\omega}_c c^\dagger c \\
&\quad + \chi b^\dagger b c^\dagger c + \frac{K_c}{2} c^{\dagger 2} c^2 + \Omega_0(b+b^\dagger),
\end{aligned}
\end{equation}
where the effective parameters are given by
\begin{align}
\bar{\omega}_b &\approx \omega_b + \frac{g^2}{\Delta_{bc}} - \frac{g^4}{\Delta_{bc}^3}, \\
\bar{\omega}_c &\approx \omega_c - \frac{g^2}{\Delta_{bc}} + \frac{g^4}{\Delta_{bc}^3}, \\
\chi &\approx \frac{2K g^2}{\Delta_{bc}^2} - \frac{2K^2 g^2}{\Delta_{bc}^3} - \frac{8K g^4}{\Delta_{bc}^4}, \\
K_c &\approx \frac{K g^4}{\Delta_{bc}^4}.
\end{align}
The small expansion parameters are $g/|\Delta_{bc}|\ll1$ and $|K/\Delta_{bc}|\ll1$. We treat the drive perturbatively and obtain the fourth-order corrections to the transition frequency:
\begin{equation}
\begin{aligned}
\widetilde{\omega}_{c;n,m} &= \bar{\omega}_c(n-m) + \frac{K_c}{2}[n(n-1) - m(m-1)] \\
&\quad + \frac{\Omega_0^2}{-\Delta_{bd}-n\chi} - \frac{\Omega_0^2}{-\Delta_{bd}-m\chi} \\
&\quad - \frac{\Omega_0^4}{(-\Delta_{bd}-n\chi)^3}+ \frac{\Omega_0^4}{(-\Delta_{bd}-m\chi)^3} \\
&\quad + \frac{2\Omega_0^4}{(-\Delta_{bd}-n\chi)^2(-2\Delta_{bd}-K-2n\chi)} \\
&\quad - \frac{2\Omega_0^4}{(-\Delta_{bd}-m\chi)^2(-2\Delta_{bd}-K-2m\chi)}.
\end{aligned}
\end{equation}
The perturbative expansion is organized in the small quantities $g$, $\Omega_0$, and $K$. In the regime of interest, $|\chi/\Delta_{bd}|\ll1$, we keep terms up to fourth order to obtain
\begin{equation}
\label{unselective frequency}
\begin{aligned}
\widetilde{\omega}_{c;n,m} &\approx (n-m) \bigg[ \omega_c -\frac{g^2}{\Delta_{bc}} +\frac{g^4}{\Delta_{bc}^3} \\
&\quad +2\frac{\Omega_0^2}{\Delta_{bd}^{2}} \frac{g^2}{\Delta_{bc}^{2}}K -\frac{K_c}{2} \bigg] + (n^2-m^2) \frac{K_c}{2},
\end{aligned}
\end{equation}
where we have expanded to leading order in $\chi/\Delta_{bd}$. There is no correction to the cavity transition frequency at order $\Omega_0^4$ alone. Because the drive acts on the FTT, it affects the cavity only through the dispersive cavity--FTT coupling.
Taking the derivative with respect to flux gives

\begin{align}
&\frac{d \widetilde{\omega}_{c;n,m}}{d \Phi} \approx (n-m) \bigg[ \frac{g^2}{\Delta_{bc}^2} - \frac{3g^4}{\Delta_{bc}^4} \\
&\quad - 4\frac{\Omega_{0}^{2} g^2 K}{\Delta_{bc}^{2}\Delta_{bd}^{2}} \bigg( \frac{1}{\Delta_{bc}} + \frac{1}{\Delta_{bd}} \bigg) + 2\frac{K g^4}{\Delta_{bc}^{5}} \bigg]\frac{\partial\omega_b}{\partial \Phi} \\
&\quad -(n^2-m^2)\bigg[ 2\frac{K g^4}{\Delta_{bc}^{5}} \bigg] \frac{\partial\omega_b}{\partial \Phi} \\
&\approx (n-m) \frac{g^2}{\Delta_{bc}^2} \bigg[ 1 - \frac{3g^2}{\Delta_{bc}^2} - \frac{4\Omega_{0}^{2} K}{\Delta_{bd}^{2}} \bigg( \frac{1}{\Delta_{bc}} + \frac{1}{\Delta_{bd}} \bigg) \\
&\quad +2\frac{K g^2}{\Delta_{bc}^{3}} \bigg] \frac{\partial\omega_b}{\partial \Phi} -(n^2-m^2) \bigg[ 2\frac{K g^4}{\Delta_{bc}^{5}} \bigg] \frac{\partial\omega_b}{\partial \Phi} \\
&\approx (n-m) \frac{g^2}{\Delta_{bc}^2} \bigg[ 1 - \frac{4\Omega_{0}^{2} }{\Delta_{bd}^2} \frac{K}{\Delta_{bd}} \bigg] \frac{\partial\omega_b}{\partial \Phi}.
\end{align}

Setting the derivative to zero reproduces the sweet-spot condition in the main text.

\subsection{Number of Protected Levels.}
To quantify the range of protected cavity levels, we compare the residual sensitivities. We define
\begin{equation}
r_{n,m} \equiv \left| \frac{d\widetilde{\omega}_{c;n,m}/d\Phi}{\left. d\widetilde{\omega}_{c;n,m}/d\Phi \right|_{\Omega_0=0}} \right|.
\end{equation}
We say that the transition is protected within a tolerance $\epsilon$ if $r_{n,m}<\epsilon$. At fourth order, the drive can cancel the linear flux susceptibility, so that $r_{n,m}=0$ at the sweet spot.
To estimate this residual sensitivity, we use a sixth-order perturbative calculation and retain terms through seventh order in the small parameters in the  cavity transition frequency. 

Its derivative with respect to $\Phi$ gives
\begin{widetext}
\begin{align}
\label{higher-order sensitivity}
\frac{d \widetilde{\omega}_{c;n,m}}{d\Phi} &\approx (n-m) \Bigg[ \frac{g^2}{\Delta_{bc}^2} -\frac{3g^4}{\Delta_{bc}^4} +2\frac{K g^4}{\Delta_{bc}^{5}} +84\frac{K^2g^6}{\Delta_{bc}^8} -4\frac{K\Omega_0^2g^2}{\Delta_{bc}^3\Delta_{bd}^2} -4\frac{K\Omega_0^2g^2}{\Delta_{bc}^2\Delta_{bd}^3} \\
&\qquad +16\frac{K^2\Omega_0^4g^2}{\Delta_{bc}^3\Delta_{bd}^4(2\Delta_{bd}+K)^3} \big( 10\Delta_{bc}\Delta_{bd} +3\Delta_{bc}K +4\Delta_{bd}^2 +2\Delta_{bd}K \big) \\
&\qquad +2\frac{K^2\Omega_0^2g^2(2\Delta_{bc}+3\Delta_{bd})}{\Delta_{bc}^4\Delta_{bd}^3} -4\frac{K^3\Omega_0^2g^2(\Delta_{bc}+2\Delta_{bd})}{\Delta_{bc}^5\Delta_{bd}^3} +16\frac{K\Omega_0^2g^4(\Delta_{bc}+2\Delta_{bd})}{\Delta_{bc}^5\Delta_{bd}^3} \\
&\qquad +2\frac{K^4\Omega_0^2g^2(2\Delta_{bc}+5\Delta_{bd})}{\Delta_{bc}^6\Delta_{bd}^3} -16\frac{K^2\Omega_0^2g^4(2\Delta_{bc}+5\Delta_{bd})}{\Delta_{bc}^6\Delta_{bd}^3} \Bigg] \frac{\partial\omega_b}{\partial\Phi} \\
&\quad +(n^2-m^2) \Bigg[ -2\frac{K g^4}{\Delta_{bc}^{5}} -126\frac{K^2g^6}{\Delta_{bc}^8} + 4\frac{K^2\Omega_0^2g^4(3\Delta_{bc}+4\Delta_{bd})}{\Delta_{bc}^{5}\Delta_{bd}^{4}} \Bigg] \frac{\partial\omega_b}{\partial\Phi} +42(n^3-m^3) \frac{K^2g^6}{\Delta_{bc}^8} \frac{\partial\omega_b}{\partial\Phi}.
\end{align}
\end{widetext}
The terms proportional to $n-m$ define the linear susceptibility. Higher-order corrections to this coefficient only perturbatively shift the sweet-spot location. We therefore define the corrected sweet spot by requiring the full coefficient of $n-m$ to vanish. The remaining terms, proportional to $n^2-m^2$ and $n^3-m^3$, cannot be removed by this single condition and yield the residual nonlinear susceptibility:
\begin{align}
&\left. \frac{d\widetilde{\omega}_{c;n,m}}{d\Phi} \right|_{\rm nl} = (n^2 - m^2) \Bigg[ -2\frac{K g^4}{\Delta_{bc}^5} \\
&\quad + 4 \frac{K^2\Omega_0^2 g^4(3\Delta_{bc} + 4\Delta_{bd})}{\Delta_{bc}^5 \Delta_{bd}^4} - 126 \frac{K^2 g^6}{\Delta_{bc}^8} \Bigg] \frac{\partial \omega_b}{\partial \Phi} \nonumber \\
&\quad + 42(n^3 - m^3) \frac{K^2 g^6}{\Delta_{bc}^8} \frac{\partial \omega_b}{\partial \Phi}. \nonumber
\end{align}
To estimate the leading residual sensitivity, it is sufficient to evaluate this expression at the leading-order sweet spot, because the shift of the corrected sweet spot produces only higher-order corrections to $\left.d\widetilde{\omega}_{c;n,m}/d\Phi\right|_{\rm nl}$. We therefore use $\Delta_{bd}^3=4K\Omega_0^2$, obtaining
\begin{align}
&\left. \frac{d \widetilde{\omega}_{c;n,m}}{d\Phi} \right|_{\rm nl} = \Bigg\{ (n^2-m^2) \Bigg[ \frac{K g^4}{\Delta_{bc}^{5}} \bigg( 2+3\frac{\Delta_{bc}}{\Delta_{bd}} \bigg)\nonumber \\
&\quad -126\frac{K^2g^6}{\Delta_{bc}^{8}} \Bigg]+ 42(n^3-m^3) \frac{K^2g^6}{\Delta_{bc}^{8}} \Bigg\} \frac{\partial\omega_b}{\partial\Phi} \nonumber \\
&\approx \Bigg[ 3(n^2-m^2) \frac{K g^4}{\Delta_{bc}^{4}\Delta_{bd}} \nonumber \\
&\quad + 42\big((n^3-m^3)-3(n^2-m^2)\big) \frac{K^2g^6}{\Delta_{bc}^{8}} \Bigg] \frac{\partial\omega_b}{\partial\Phi}.
\end{align}
where, in the last step, we used $|\Delta_{bd}|\ll|\Delta_{bc}|$. In the low-photon regime where the static cubic Kerr contribution remains subleading, the dominant residual is therefore
\begin{equation}
\left. \frac{d \widetilde{\omega}_{c;n,m}}{d\Phi} \right|_{\rm nl} \approx 3(n^2-m^2) \frac{K g^4}{\Delta_{bc}^{4}\Delta_{bd}} \frac{\partial\omega_b}{\partial\Phi}.
\end{equation}
The undriven leading-order susceptibility is
\begin{equation}
\left. \frac{d\widetilde{\omega}_{c;n,m}}{d\Phi} \right|_{\Omega_0=0} \approx (n-m)\frac{g^2}{\Delta_{bc}^2} \frac{\partial\omega_b}{\partial\Phi}.
\end{equation}
Thus, for $n\neq m$, we obtain
\begin{equation}
\begin{aligned}
\epsilon_{n,m} &\approx \left| \frac{ 3(n^2-m^2)K g^4/(\Delta_{bc}^{4}\Delta_{bd}) }{ (n-m)g^2/\Delta_{bc}^2 } \right| \\
&= 3|n+m| \left| \frac{K g^2}{\Delta_{bc}^{2}\Delta_{bd}} \right|.
\end{aligned}
\end{equation}
Using the leading-order dispersive shift
\begin{equation}
\chi \approx \frac{2K g^2}{\Delta_{bc}^{2}},
\end{equation}
we obtain
\begin{equation}
\epsilon_{n,m} \approx \frac{3}{2}|n+m| \left| \frac{\chi}{\Delta_{bd}} \right|.
\end{equation}
Requiring $\epsilon_{n,m}<\epsilon$ gives
\begin{equation}
|n+m| < \frac{\epsilon}{3} \left| \frac{\Delta_{bc}^{2}\Delta_{bd}}{K g^2} \right| = \frac{2\epsilon}{3} \left| \frac{\Delta_{bd}}{\chi} \right|.
\end{equation}
If we aim to protect all transitions among cavity levels $0\le m<n\le N$, the largest value of $n+m$ is $2N-1$. Therefore,
\begin{equation}
2N-1 < \frac{2\epsilon}{3} \left| \frac{\Delta_{bd}}{\chi} \right| \quad \Rightarrow \quad N < \frac{\epsilon}{3} \left| \frac{\Delta_{bd}}{\chi} \right|+\frac{1}{2}.
\end{equation}
This gives the estimate shown in the main text,
\begin{equation}
N = \max \left( 1, \left\lfloor \frac{\epsilon}{3} \left| \frac{\Delta_{bd}}{\chi} \right| +\frac{1}{2} \right\rfloor \right).
\end{equation}
\subsection{Multimode Sweet-Spot Alignment}
\label{Multimode Sweet-Spot Alignment}
We now ask whether a single driven FTT can simultaneously protect multiple cavity modes. In practice, for a fixed drive amplitude $\Omega_0$, one can sweep the drive detuning $\Delta_{bd}$ and evaluate the flux susceptibility of each cavity transition. The minima of the susceptibilities for different cavity modes mightnot occur at exactly the same drive detuning. In this subsection, we estimate analytically what determines the spread of these minima.

We assume that the cavity modes coupled to the same FTT are mutually far detuned, so that FTT-mediated cavity--cavity exchange is suppressed. We also choose the cavity frequencies such that residual cavity--cavity cross-Kerr interactions are negligible~\cite{Zhu2013}. After neglecting these residual intermode couplings, the rotating-frame Hamiltonian is
\begin{equation}
\begin{aligned}
H_{R}^{\mathrm{multi}}
&= \Delta_{bd} b^\dagger b + \frac{K}{2} b^{\dagger 2} b^2 + \Omega_0(b+b^\dagger) \\
&\quad + \sum_j \bigg[
\bar{\omega}_{c,j} c_j^\dagger c_j + \chi_j b^\dagger b c_j^\dagger c_j + \frac{K_{c,j}}{2} c_j^{\dagger 2} c_j^2
\bigg],
\end{aligned}
\end{equation}
where $c_j$ is the annihilation operator for cavity mode $j$. For each cavity mode, the leading-order flux susceptibility has the form
\begin{equation}
\frac{d\widetilde{\omega}_{c_j}}{d\Phi}
\approx
\frac{g_j^2}{\Delta_{bc,j}^2}
\left[
1-\frac{4K\Omega_0^2}{\Delta_{bd}^3}
\right]
\frac{\partial\omega_b}{\partial\Phi}.
\end{equation}
Therefore, at leading order, all cavity modes are protected at the same detuning,
\begin{equation}
\Delta_{bd,0}^3=4K\Omega_0^2 .
\end{equation}
The spread of the minima appears only beyond this leading order. Thus, the number of cavity modes that can be protected by a single driven FTT is controlled by how closely their higher-order corrected sweet-spot detunings remain aligned.

To estimate this spread, we use Eq.~\eqref{higher-order sensitivity} and replace $g\rightarrow g_j$ and $\Delta_{bc}\rightarrow\Delta_{bc,j}$ for each cavity mode. The relevant quantity is the coefficient of $n-m$, since it determines the linear flux susceptibility of mode $j$. Keeping the leading mode-dependent corrections, this coefficient can be written as
\begin{equation}
\begin{aligned}
\frac{d\widetilde{\omega}_{c,j}}{d\Phi}
&\approx
\frac{g_j^2}{\Delta_{bc,j}^2}
Q_j(\Delta_{bd})
\frac{\partial\omega_b}{\partial\Phi}, \\
Q_j(\Delta_{bd})
&=
1-\frac{4K\Omega_0^2}{\Delta_{bd}^3}
-\frac{4K\Omega_0^2}{\Delta_{bc,j}\Delta_{bd}^2}
-3\frac{g_j^2}{\Delta_{bc,j}^2}
+\cdots .
\end{aligned}
\end{equation}
The first two terms give the leading common sweet spot, while the remaining terms shift the location of the minimum in a mode-dependent way. Let
\begin{equation}
\Delta_{bd,j}^*=\Delta_{bd,0}+\delta\Delta_j,
\qquad
\Delta_{bd,0}^3=4K\Omega_0^2 ,
\end{equation}
where $\Delta_{bd,j}^*$ is the corrected minimum for cavity mode $j$. Expanding $Q_j(\Delta_{bd,j}^*)=0$ around $\Delta_{bd,0}$, we obtain
\begin{equation}
Q_j(\Delta_{bd,0})
+
\left.
\frac{d}{d\Delta_{bd}}
\left(
1-\frac{4K\Omega_0^2}{\Delta_{bd}^3}
\right)
\right|_{\Delta_{bd,0}}
\delta\Delta_j \approx0.
\end{equation}
Since
\begin{equation}
\left.
\frac{d}{d\Delta_{bd}}
\left(
1-\frac{4K\Omega_0^2}{\Delta_{bd}^3}
\right)
\right|_{\Delta_{bd,0}}
= \frac{3}{\Delta_{bd,0}},
\end{equation}
the mode-dependent shift is
\begin{equation}
\delta\Delta_j
\approx
\frac{\Delta_{bd,0}^2}{3\Delta_{bc,j}}
+
\Delta_{bd,0}\frac{g_j^2}{\Delta_{bc,j}^2}
- \frac{2}{3}\Delta_{bd,0}\frac{K g_j^2}{\Delta_{bc,j}^3}
+\cdots .
\end{equation}
Therefore, the spread between the minima of two cavity modes $i$ and $j$ at leading order is approximately
\begin{align}
   \Delta_{bd,i}-\Delta_{bd,j} &\approx \frac{\Delta_{bd,0}^2}{3} \left( \frac{1}{\Delta_{bc,i}} - \frac{1}{\Delta_{bc,j}} \right)\nonumber\\
    &=\frac{(4|K|\Omega_0^2)^{2/3}}{3}
\left(
\frac{1}{\Delta_{bc,i}}
-
\frac{1}{\Delta_{bc,j}}
\right).
\end{align}

Thus, the leading spread is controlled by the variation of $1/\Delta_{bc,j}$ across the cavity modes. In the ideal unselective limit, where $|\Delta_{bd,0}/\Delta_{bc,j}|\ll1$, these shifts are small and the susceptibility minima remain aligned.

\section{Example of a SNAIL Coupler}
\label{snail}

A SNAIL is a three-wave-mixing circuit element widely used for parametric amplification and gate operations \cite{SNAIL}.
When a SNAIL is capacitively coupled to a cavity, the system Hamiltonian can be written as
\begin{equation}
\begin{aligned}
H =\;& \omega_s s^\dagger s 
+ \sum_j g_j (s + s^\dagger)^j
+ \omega_c c^\dagger c \\
&+ g (s + s^\dagger)(c + c^\dagger)
+ 2\Omega_0 \cos(\omega_d t) (s + s^\dagger).
\end{aligned}
\end{equation}
Here \(\omega_s\) is the SNAIL frequency, \(g_j\) are the nonlinear coefficients, \(g\) is the SNAIL--cavity coupling strength, \(s\) (\(c\)) is the SNAIL (cavity) mode operator, and \(\omega_d\) is the drive frequency.
The parameters \(\omega_s\), \(g_j\), \(g\), and \(\Omega_0\) are flux dependent.

\begin{figure}[!htbp]
    \centering
    \includegraphics[width=0.9\linewidth]{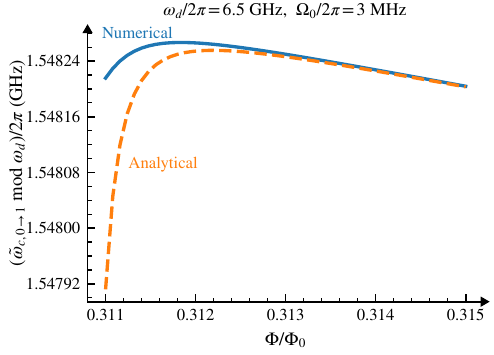}
    \caption{Dynamical sweet spot in a SNAIL--cavity system. Under continuous drive, the cavity transition energy exhibits a maximum near \(\Phi \approx 0.311\,\Phi_0\), indicating a flux sweet spot. The solid lines show numerical results, and the dashed lines show the perturbative prediction from Eq.~\eqref{energy correction}.}
    \label{fig:snail_ss}
\end{figure}

We apply SAFE to generate a flux sweet spot in the cavity frequency and choose a specific set of system parameters for the illustration. The SNAIL is biased at \(\Phi = 0.3\,\Phi_0\), where the cubic nonlinearity \(g_3\) enables three-wave mixing.
At this operating point, \(g_3/2\pi \approx 67~\mathrm{MHz}\), \(\omega_s/2\pi = 6.59~\mathrm{GHz}\), and \(\omega_c/2\pi = 5~\mathrm{GHz}\). A coherent drive with amplitude \(\Omega_0 = 3~\mathrm{MHz}\) at \(\omega_d = 6.5~\mathrm{GHz}\) produces a flux sweet spot in the cavity spectrum, as shown in Fig.~\ref{fig:snail_ss}. The cavity transition energy has a local maximum near \(\Phi \approx 0.311\,\Phi_0\), which is captured by the perturbative expression in Eq.~\eqref{energy correction}.

\section{Dephasing Induced by Drive-Amplitude Noise}
\label{app:amplitude_noise}
We can tune the drive parameters so that the shifted cavity frequency
\begin{equation}
\widetilde{\omega}_{c,0\rightarrow 1}=\bar{\omega}_c+\frac{\Omega_0^2\chi}{\Delta_{bd}^2}
\end{equation}
is insensitive to flux noise to leading order. However, this frequency is sensitive to drive-amplitude noise. 

We consider a spectrum of drive-amplitude noise containing both white-noise and $1/f$-noise components,
\begin{equation}
    S_\Omega(\omega) = (A^\mathrm{w})^2 + \frac{(A^{1/f})^2}{|\omega|/2\pi},
\end{equation}
where $A^\mathrm{w}$ and $A^{1/f}$ denote the amplitudes of the white-noise and $1/f$-noise contributions, respectively. The corresponding cavity dephasing rate is~\cite{PhysRevApplied.12.054015dss2}
\begin{equation}
    \frac{1}{T_\phi}
    =
    (A^\mathrm{w})^2
    \bigg|\frac{\partial \widetilde{\omega}_{c,0\rightarrow 1}}{\partial \Omega}\bigg|^2
    +
    A^{1/f}
    \bigg|\frac{\partial \widetilde{\omega}_{c,0\rightarrow 1}}{\partial \Omega}\bigg|
    \sqrt{2|\ln(\omega_\text{ir} t)|}.
\end{equation}

We evaluate this expression for a specific hardware platform to estimate the corresponding dephasing time. For the Zurich Instruments HDAWG~\cite{zhinstHDAWGmanual}, the white-noise floor is $25\,\mathrm{nV}/\sqrt{\mathrm{Hz}}$, and the $1/f$ corner frequency is approximately $100\,\mathrm{kHz}$. These specifications imply a $1/f$ voltage-noise spectral density of approximately $7.9\,\mu\mathrm{V}/\sqrt{\mathrm{Hz}}$ at $1\,\mathrm{Hz}$.

To convert the voltage noise into drive-amplitude noise, we use the linear relation between the drive voltage $V$ and the induced Rabi frequency $\Omega$. Cryogenic attenuation suppresses the signal and the drive-amplitude noise by the same factor, so that $\delta\Omega/\Omega = \delta V/V$. The corresponding white-noise and $1/f$-noise amplitudes of the drive are therefore
\begin{equation}
    A^\mathrm{w} = \Omega \times \frac{25\,\mathrm{nV}/\sqrt{\mathrm{Hz}}}{V_{\mathrm{AWG}}},
    \qquad
    A^{1/f} = \Omega \times \frac{7.9\,\mu\mathrm{V}}{V_{\mathrm{AWG}}},
\end{equation}
where $V_{\mathrm{AWG}}$ is the peak output voltage of the instrument. For illustration, we choose a target drive strength of $\Omega/2\pi = 10\,\mathrm{MHz}$ and $V_{\mathrm{AWG}}=0.5\,\mathrm{V}$. This gives $A^\mathrm{w} \approx 0.5\,\sqrt{\mathrm{Hz}}$ and $A^{1/f} \approx 158\,\mathrm{Hz}$.

Using the parameters at the sweet spot in Fig.~\ref{fig:compare}(a), we numerically evaluate
\begin{equation}
    \bigg|\frac{\partial \widetilde{\omega}_{c,0\rightarrow 1}}{\partial \Omega}\bigg| = 0.044.
\end{equation}
Assuming a typical measurement timescale such that $\sqrt{|\ln(\omega_\text{ir} t)|} \approx 4$~\cite{PhysRevX.7.031037factor4}, we obtain a pure-dephasing rate of
\begin{equation}
    \frac{1}{T_\phi} \approx 39\,\mathrm{s}^{-1},
\end{equation}
which corresponds to a dephasing time limited by drive-amplitude noise of
\begin{equation}
    T_\phi \approx 25\,\mathrm{ms}.
\end{equation}

\section{Exact Calculation of the Sensitivity of the Cavity Frequency}
\label{app:numerics}
This appendix specifies system parameters and numerical details used to obtain the results in Sec.\ \ref{sec3} and \ref{sec4}.
The driven circuit is simulated using the full system Hamiltonian $H_{\mathrm{sys}}$ in Eq.~\eqref{Hfull}. We use symmetric junctions with $E_{J1}=E_{J2}=30.19\,\mathrm{GHz}$, so that $E_J=60.38\,\mathrm{GHz}$. Other parameters are $E_C=0.10\,\mathrm{GHz}$, $\omega_c/2\pi=5.226\,\mathrm{GHz}$, $g'/2\pi=0.050\,\mathrm{GHz}$, and $\Phi=0.20\,\Phi_0$.
For these values, the effective parameters appearing in Eq.~\eqref{Hfull} are $\omega_b/2\pi=6.16\,\mathrm{GHz}$, $K/2\pi=-0.10\,\mathrm{GHz}$, and $g/2\pi=0.10\,\mathrm{GHz}$.

The basis is obtained by diagonalizing the static part of $H_{\mathrm{sys}}$, that is, $H_{\mathrm{sys}}(t)$ with $\Omega_0=\delta\Phi(t)=0$.
In the dispersive regime, these eigenstates are labeled by the corresponding bare basis states.
The FTT and cavity Hilbert spaces are truncated to three and four levels, respectively. Increasing the truncation changes the results only marginally and does not affect the conclusions.

Floquet simulations are performed directly using $H_{\mathrm{sys}}(t)$ to obtain time-periodic Floquet modes $\ket{\phi_\alpha(t)}$ and quasienergies $\epsilon_\alpha$, with $\ket{\phi_\alpha(t+T)}=\ket{\phi_\alpha(t)}$ and $T=2\pi/\omega_d$.
To assign each Floquet mode a label in a chosen bare basis $\{\ket{j}\}$, the time-averaged overlap weight
\begin{equation}
W_{\alpha j}
= \frac{1}{T} \int_{0}^{T} dt \,
\left| \langle j \mid \phi_\alpha(t) \rangle \right|^2
\end{equation}
is evaluated.
Each Floquet mode $\alpha$ is labeled by the basis state $\ket{j}$ with the largest weight,
\begin{equation}
\mathrm{label}(\alpha)=\arg\max_j W_{\alpha j}.
\end{equation}
This procedure yields a relabeling $\alpha\mapsto(n,m)$, where $n$ and $m$ denote the FTT and cavity excitation numbers, respectively, together with the corresponding quasienergies $\epsilon_{nm}$.

Specifically, we compute
\begin{equation}
D_{0\rightarrow1}=\frac{\partial}{\partial\Phi}\!\left(\epsilon_{g1}-\epsilon_{g0}\right),
\end{equation}
which yields the exact $D_{0\rightarrow1}$ reported in Fig.~\ref{fig:compare}.
\section{Effective Lindblad Master Equation}
This section derives the effective Lindblad master equation for Eq.~\eqref{Lindblad}.
In the dispersive frame, the equation becomes~\cite{cqedreview2021}
\begin{align}
\dot{\rho}_{\mathrm{disp}}
&= -i\left[H_{\mathrm{disp}}+H_{\text{drive}}(t)+H_{\text{noise,disp}}(\delta\Phi(t)), \rho_{\mathrm{disp}}\right] \notag \\
&\quad + \left[\gamma_{c\downarrow} + \left(\frac{g}{\Delta_{bc}}\right)^2 \gamma_{b\downarrow}\right]\mathcal{D}[c]\rho_{\mathrm{disp}} \notag \\
&\quad + \left[\gamma_{b\downarrow} + \left(\frac{g}{\Delta_{bc}}\right)^2 \gamma_{c\downarrow}\right]\mathcal{D}[b]\rho_{\mathrm{disp}} \notag \\
&\quad + \left[\gamma_{c\uparrow} + \left(\frac{g}{\Delta_{bc}}\right)^2 \gamma_{b\uparrow}\right]\mathcal{D}[c^\dagger]\rho_{\mathrm{disp}} \notag \\
&\quad + \left[\gamma_{b\uparrow} + \left(\frac{g}{\Delta_{bc}}\right)^2 \gamma_{c\uparrow}\right]\mathcal{D}[b^\dagger]\rho_{\mathrm{disp}},
\end{align}
where
\begin{equation}
H_{\text{noise,disp}} = \delta \omega_b(t) \left( b^\dagger b + \frac{g^2}{\Delta_{bc}^2} c^\dagger c - \frac{4Kg^2}{\Delta_{bc}^3} b^\dagger b c^\dagger c \right).
\end{equation}
Here, the dissipators have been transformed to the dispersive frame, giving the usual Purcell and inverse-Purcell corrections to the relaxation and excitation rates. The noise Hamiltonian describes how flux noise of the FTT also induces fluctuations of the cavity frequency and the dispersive shift, with strengths proportional to $g^2/\Delta_{bc}^2$ and $-4K g^2/\Delta_{bc}^3$, respectively. We neglect the Purcell contribution to FTT relaxation because cavity lifetimes are typically on the order of tens of milliseconds, much longer than the tens-of-microseconds relaxation times of FTTs. The master equation therefore simplifies to
\begin{align}
\dot{\rho}_{\mathrm{disp}} &= -i\left[H_{\mathrm{disp}}+H_{\text{drive}}(t)+H_{\text{noise,disp}}(\delta\Phi(t)), \rho_{\mathrm{disp}}\right] \nonumber \\
&\quad + \left[\gamma_{c\downarrow}+\left(\frac{g}{\Delta_{bc}}\right)^2 \gamma_{b\downarrow}\right]\mathcal{D}[c]\rho_{\mathrm{disp}} + \gamma_{b\downarrow}\mathcal{D}[b]\rho_{\mathrm{disp}} \nonumber \\
&\quad + \left[\gamma_{c\uparrow} + \left(\frac{g}{\Delta_{bc}}\right)^2 \gamma_{b\uparrow}\right]\mathcal{D}[c^\dagger]\rho_{\mathrm{disp}} + \gamma_{b\uparrow}\mathcal{D}[b^\dagger]\rho_{\mathrm{disp}}.
\end{align}
In the frame that corotates with the drive, we obtain the master equation
\begin{align}
\dot{\rho}_R &= -i\left[H_R+H_{\mathrm{noise,disp}}(\delta\Phi(t)), \rho_R\right] \nonumber \\
&\quad + \left[\gamma_{c\downarrow}+\left(\frac{g}{\Delta_{bc}}\right)^2 \gamma_{b\downarrow}\right]\mathcal{D}[c]\rho_R + \gamma_{b\downarrow}\mathcal{D}[b]\rho_R \nonumber \\
&\quad + \left[\gamma_{c\uparrow} + \left(\frac{g}{\Delta_{bc}}\right)^2 \gamma_{b\uparrow}\right]\mathcal{D}[c^\dagger]\rho_R + \gamma_{b\uparrow}\mathcal{D}[b^\dagger]\rho_R,
\end{align}
with $\rho_R = U_R^\dagger(t)\rho_{\mathrm{disp}}U_R(t)$. In the following, we first apply a Schrieffer--Wolff transformation (SWT) to the dispersive Hamiltonian to obtain the effective driven Hamiltonian and then transform the full Lindblad master equation.

\subsection{Effective Drive Hamiltonian}
We rewrite $H_{\mathrm{disp}}$ in terms of the number operators $n_b = b^\dagger b$ and $n_c = c^\dagger c$. Using the relation $b^{\dagger 2} b^2 = b^\dagger b (b^\dagger b - 1) = n_b(n_b - 1)$, we can express the unperturbed Hamiltonian $H_0$ and the perturbation $V$ as
\begin{align}
H_0 &= \Delta_{bd} n_b + \frac{K}{2} n_b(n_b - 1) + \overline{\omega}_c n_c + \chi n_b n_c, \\
V &= \Omega_0 (b + b^\dagger).
\end{align}
For a cleaner derivation, we define an operator $\mathcal{E}(n_b)$ that represents the energy difference between the $\ket{n_b}$ and $\ket{n_b - 1}$ states:
\begin{equation}
\mathcal{E}(n_b) \equiv \Delta_{bd} + K n_b + \chi n_c .
\end{equation}
The first-order generator in the SWT is chosen to eliminate the perturbation to first order, which requires solving
\begin{equation}
[D_{d1}, H_0] = -V.
\end{equation}
We obtain
\begin{equation}
D_{d1} = \Omega_0 \left( b^\dagger \frac{1}{\mathcal{E}(n_b)} - \frac{1}{\mathcal{E}(n_b)} b \right).
\end{equation}
The second-order effective Hamiltonian is generated by
\begin{equation}
H^{(2)} = \frac{1}{2} [D_{d1}, V].
\end{equation}
Substituting $D_{d1}$ and $V$ gives
\begin{equation}
H^{(2)} = \frac{\Omega_0^2}{2} \left[ b^\dagger \mathcal{E}(n_b)^{-1} - \mathcal{E}(n_b)^{-1} b, b + b^\dagger \right].
\end{equation}
The diagonal term of the second-order correction is
\begin{equation}
H^{(2)}_{\text{diag}} = \Omega_0^2 \left( \frac{n_b}{\mathcal{E}(n_b - 1)} - \frac{n_b + 1}{\mathcal{E}(n_b)} \right).
\end{equation}
Expanding this correction to fourth order in the small parameters, subtracting the ground-state energy, and keeping the unperturbed Hamiltonian, we obtain the effective Hamiltonian
\begin{align}
H_d &= \overline{\omega}_c n_c + \Delta_{bd} n_b + \frac{K}{2} n_b(n_b - 1)+ \chi n_b n_c \nonumber \\
&\quad + \Omega_0^2 \left( \frac{n_b}{\Delta_{bd} + K(n_b - 1)} - \frac{n_b + 1}{\Delta_{bd} + K n_b} + \frac{1}{\Delta_{bd}} \right) \nonumber \\
&\quad - \Omega_0^2 \chi n_c \left( \frac{n_b}{(\Delta_{bd} + K(n_b - 1))^2} - \frac{n_b + 1}{(\Delta_{bd} + K n_b)^2} \right).
\end{align}
Taking the FTT to be in its ground state reproduces the expression for the cavity transition frequencies $\tilde{\omega}_{c;n,m}$.

\subsection{Transformation of the Dissipators and the Noise Hamiltonian}
\label{Transformation of the Dissipators and the Noise Hamiltonian}
In this section, we transform the dissipators and the noise Hamiltonian into the frame of $H_d$. The relevant operators are $b$, $c$, $b^\dagger b$, and $c^\dagger c$. The transformation of an operator can be approximated using the Baker--Campbell--Hausdorff (BCH) series,
\begin{equation}
O' = O + [D_{d1}, O] + [D_{d2}, O] + \frac{1}{2}[D_{d1}, [D_{d1}, O]],
\end{equation}
where the second-order generator obtained from the SWT is
\begin{equation}
D_{d2} = -\frac{\Omega_0^2 K}{2} \Big( b^{\dagger 2} F(n_b) - F(n_b) b^2 \Big),
\end{equation}
with
\begin{equation}
F(n_b) = \frac{1}{\mathcal{E}(n_b)\mathcal{E}(n_b+1)\big(\mathcal{E}(n_b) + \mathcal{E}(n_b+1)\big)}.
\end{equation}

\subsubsection{Transformation of the FTT Annihilation Operator}
Summing the respective commutators and grouping terms by operator order, namely the constant, $b$, $b^\dagger$, $b^2$, and $b^3$ terms, yields the dressed annihilation operator
\begin{align}
b' &= \Omega_0 \left( \frac{n_b}{\mathcal{E}(n_b - 1)} + \frac{n_b + 1}{\mathcal{E}(n_b)} \right) \nonumber \\
&\quad + \left( 1 + \frac{\Omega_0^2}{2} C_b(n_b) \right) b \nonumber \\
&\quad + b^\dagger \left[ \frac{\Omega_0^2}{2} C_{b^\dagger}(n_b) - \frac{\Omega_0^2 K}{2} \Big( n_b F(n_b - 1) - (n_b + 2) F(n_b) \Big) \right] \nonumber \\
&\quad - \Omega_0 \left( \frac{1}{\mathcal{E}(n_b)} + \frac{1}{\mathcal{E}(n_b + 1)} \right) b^2 \nonumber \\
&\quad + \frac{\Omega_0^2 K}{2} \Big( F(n_b) - F(n_b + 1) \Big) b^3,
\end{align}
where
\begin{align}
C_b(n_b) &= n_b \left( \frac{1}{\mathcal{E}(n_b)} + \frac{1}{\mathcal{E}(n_b-1)} \right)^2 \nonumber \\
&\quad - (n_b+2) \left( \frac{1}{\mathcal{E}(n_b)} + \frac{1}{\mathcal{E}(n_b+1)} \right)^2, \\
C_{b^\dagger}(n_b) &= \frac{1}{\mathcal{E}(n_b)} \left( \frac{n_b}{\mathcal{E}(n_b-1)} + \frac{2(n_b+1)}{\mathcal{E}(n_b)} + \frac{n_b+2}{\mathcal{E}(n_b+1)} \right).
\end{align}

\subsubsection{Transformation of the FTT Number Operator}
Applying the BCH series to $n_b$ yields a mixture of diagonal terms, linear drive terms, and two-photon terms:
\begin{align}
n_b' &= n_b + \Omega_0^2 \left( \frac{n_b + 1}{\mathcal{E}(n_b)^2} - \frac{n_b}{\mathcal{E}(n_b - 1)^2} \right) \nonumber \\
&\quad - \Omega_0 \left( b^\dagger \mathcal{E}(n_b)^{-1} + \mathcal{E}(n_b)^{-1} b \right) \nonumber \\
&\quad + \Omega_0^2 K \Big( b^{\dagger 2} F(n_b) + F(n_b) b^2 \Big).
\end{align}

\subsubsection{Transformation of the Cavity Annihilation Operator}
For mode $c$, all resulting terms remain proportional to $c$ because the transformations act primarily on the $b$-mode subspace while acquiring $n_c$ dependence:
\begin{align}
c' &= \left[ 1 + \frac{\Omega_0^2}{2} C_{diag}(n_b, n_c) \right] c \nonumber \\
&\quad + \chi \Omega_0 \bigg( b^\dagger \frac{1}{\mathcal{E}(n_b, n_c)\mathcal{E}(n_b, n_c + 1)} \\
&- \frac{1}{\mathcal{E}(n_b, n_c)\mathcal{E}(n_b, n_c + 1)} b \bigg) c \nonumber \\
&\quad + b^{\dagger 2} \left[ \frac{\Omega_0^2}{2} C_{b^{\dagger 2}}(n_b, n_c) - \frac{\Omega_0^2 K}{2} \Delta F(n_b, n_c) \right] c \nonumber \\
&\quad + \left[ \frac{\Omega_0^2}{2} C_{b^2}(n_b, n_c) + \frac{\Omega_0^2 K}{2} \Delta F(n_b, n_c) \right] b^2 c,
\end{align}
with the operators
\begin{align}
C_{diag}(n_b, n_c) &= - n_b \bigg( \frac{1}{\mathcal{E}(n_b-1, n_c)} + \frac{1}{\mathcal{E}(n_b-1, n_c+1)} \bigg)^2 \nonumber \\
&\quad + (n_b+1) \bigg( \frac{1}{\mathcal{E}(n_b, n_c)} + \frac{1}{\mathcal{E}(n_b, n_c+1)} \bigg)^2, \\
C_{b^{\dagger 2}}(n_b, n_c) &= \frac{1}{\mathcal{E}(n_b+1, n_c)\mathcal{E}(n_b, n_c)} \nonumber \\
&\quad + \frac{2}{\mathcal{E}(n_b+1, n_c)\mathcal{E}(n_b, n_c+1)} \nonumber \\
&\quad + \frac{1}{\mathcal{E}(n_b+1, n_c+1)\mathcal{E}(n_b, n_c+1)}, \\
C_{b^2}(n_b, n_c) &= \frac{1}{\mathcal{E}(n_b, n_c)\mathcal{E}(n_b+1, n_c)} \nonumber \\
&\quad + \frac{2}{\mathcal{E}(n_b, n_c)\mathcal{E}(n_b+1, n_c+1)} \nonumber \\
&\quad + \frac{1}{\mathcal{E}(n_b, n_c+1)\mathcal{E}(n_b+1, n_c+1)}, \\
\Delta F(n_b, n_c) &\equiv F(n_b, n_c) - F(n_b, n_c + 1).
\end{align}
The transformed master equation is
\begin{align}
\label{fullLME}
\dot{\rho}_{d} &= -i\left[H_d + H_{\text{noise},d}, \rho_{d}\right] \nonumber \\
&\quad + \left[\gamma_{c\downarrow}+\left(\frac{g}{\Delta_{bc}}\right)^2 \gamma_{b\downarrow}\right]\mathcal{D}[c']\rho_{d} + \gamma_{b\downarrow}\mathcal{D}[b']\rho_{d} \nonumber \\
&\quad + \left[\gamma_{c\uparrow} + \left(\frac{g}{\Delta_{bc}}\right)^2 \gamma_{b\uparrow}\right]\mathcal{D}[c'^{\dagger}]\rho_{d} + \gamma_{b\uparrow}\mathcal{D}[b'^\dagger]\rho_{d},
\end{align}
where the transformed noise Hamiltonian is
\begin{equation}
H_{\text{noise},d} = \delta \omega_b(t) \left( n_b' + \frac{g^2}{\Delta_{bc}^2} c^\dagger c - \frac{4Kg^2}{\Delta_{bc}^3} n_b' c^\dagger c \right).
\end{equation}
Note that $c^\dagger c$ remains unchanged because it commutes with the generators.

\subsection{Effective Lindblad Master Equation in a Truncated Subspace}
\label{Effective Lindblad Master Equation in a Truncated Subspace}
To proceed, we truncate the FTT to a two-level system, $n_b \in \{0, 1\}$. This approximation is physically justified by three conditions. First, to minimize decoherence channels acting on the logical subspace, the FTT is initialized in its ground state during idle periods. Second, the applied drive is weak and highly detuned from the FTT transition, $\Omega_0 \ll \Delta_{bd}$, so that the FTT has little excited-state population. Finally, at typical dilution-refrigerator temperatures, $T \approx 10\text{-}20$~mK, the thermal energy is much smaller than the mode frequency, $k_B T \ll \hbar \omega_b$. Thermal occupation follows a Boltzmann distribution, rendering the probability of climbing the ladder to higher excited states, $n_b \ge 2$, vanishingly small.


After the truncation, we work under the conditions $\chi \ll |\Delta_{bd}| \ll |K|$ and $\Omega_0 \ll |\Delta_{bd}|$. We keep terms up to second order in the small parameters and retain the higher-order term proportional to $\Omega_0^2\chi/\Delta_{bd}^3$, which is important for the sweet spot. The effective Hamiltonian is
\begin{align}
&H_d \approx \left( \overline{\omega}_c + \chi \frac{\Omega_0^2}{\Delta_{bd}^2} \right) n_c + \frac{K_c}{2} n_c(n_c - 1) \nonumber \\
&\quad + \left( \Delta_{bd} + \frac{2\Omega_0^2}{\Delta_{bd}} \right) \sigma_+\sigma_- \nonumber  + \chi \left( 1 - \frac{2\Omega_0^2}{\Delta_{bd}^2} \right) n_c \sigma_+\sigma_- .
\end{align}
The operators entering the noise Hamiltonian and dissipators are
\begin{align}
\sigma_-' &\approx \frac{\Omega_0}{\Delta_{bd}} \left( 1 - \frac{\chi n_c}{\Delta_{bd}} \right) (2\sigma_+\sigma_- - \mathbf{1}) \nonumber \\
&\quad + \left[ 1 - \left(\frac{\Omega_0}{\Delta_{bd}}\right)^2 \left( 1 - \frac{2\chi n_c}{\Delta_{bd}} \right) \right]\sigma_- \nonumber \\
&\quad - \left(\frac{\Omega_0}{\Delta_{bd}}\right)^2 \left( 1 - \frac{2\chi n_c}{\Delta_{bd}} \right)\sigma_+ , \\
n_b' &\approx \sigma_+ \sigma_- + \left(\frac{\Omega_0}{\Delta_{bd}}\right)^2 \left( 1 - \frac{2\chi n_c}{\Delta_{bd}} \right) (\mathbb{I} - 2\sigma_+ \sigma_-) \nonumber \\
&\quad - \frac{\Omega_0}{\Delta_{bd}} \left( 1 - \frac{\chi n_c}{\Delta_{bd}} \right) \sigma_x , \\
c' &\approx c + \frac{\chi \Omega_0}{\Delta_{bd}^2} (\sigma_+ - \sigma_-) c .
\end{align}
We then substitute these operators back into the Lindblad master equation. For example, the transformed transmon relaxation dissipator can be decomposed as
\begin{equation}
\sigma_-' = L_\downarrow + L_\uparrow + L_\phi ,
\end{equation}
where
\begin{align}
L_\downarrow &= \left[ 1 - \left(\frac{\Omega_0}{\Delta_{bd}}\right)^2 \left( 1 - \frac{2\chi n_c}{\Delta_{bd}} \right) \right]\sigma_- , \nonumber \\
L_\uparrow &= -\left(\frac{\Omega_0}{\Delta_{bd}}\right)^2 \left( 1 - \frac{2\chi n_c}{\Delta_{bd}} \right)\sigma_+ , \nonumber \\
L_\phi &= \frac{\Omega_0}{\Delta_{bd}} \left( 1 - \frac{\chi n_c}{\Delta_{bd}} \right) \left(2\sigma_+\sigma_- - \mathbf{1}\right).
\end{align}
Here, $L_\downarrow$, $L_\uparrow$, and $L_\phi$ correspond to dressed transmon relaxation, drive-induced excitation, and drive-induced dephasing, respectively. The corresponding dissipator contains both diagonal terms and cross terms,
\begin{equation}
\mathcal{D}[\sigma_-']\rho_d = \sum_{\alpha} \mathcal{D}[L_\alpha]\rho_d + \sum_{\alpha\neq\beta} \left( L_\alpha\rho_d L_\beta^\dagger - \frac{1}{2}\{L_\beta^\dagger L_\alpha, \rho_d\} \right),
\end{equation}
with $\alpha,\beta\in\{\downarrow,\uparrow,\phi\}$. In the interaction picture, these components rotate at different frequencies: $L_\phi$ is stationary, whereas $L_\downarrow$ and $L_\uparrow$ rotate near the transmon transition frequency with opposite signs. Therefore, the cross terms oscillate at frequency differences of order $\Delta_{bd}$ or $2\Delta_{bd}$. Under the secular approximation, these rapidly oscillating terms average to zero when
\begin{equation}
\gamma_{b\downarrow}\ll |\omega_\alpha-\omega_\beta| .
\end{equation}
Thus,
\begin{equation}
\mathcal{D}[\sigma_-']\rho_d \approx \mathcal{D}[L_\downarrow]\rho_d + \mathcal{D}[L_\uparrow]\rho_d + \mathcal{D}[L_\phi]\rho_d .
\end{equation}
Since $\mathcal{D}[-L]\rho_d=\mathcal{D}[L]\rho_d$, the sign of $L_\uparrow$ can be dropped. Including the bare transmon relaxation rate gives
\begin{align}
\gamma_{b\downarrow}\mathcal{D}[\sigma_-']\rho_d &\approx \gamma_{b\downarrow} \mathcal{D} \left[ \left( 1 - \left(\frac{\Omega_0}{\Delta_{bd}}\right)^2 \left( 1 - \frac{2\chi n_c}{\Delta_{bd}} \right) \right)\sigma_- \right]\rho_d \nonumber \\
&\quad + \gamma_{b\downarrow} \mathcal{D} \left[ \left(\frac{\Omega_0}{\Delta_{bd}}\right)^2 \left( 1 - \frac{2\chi n_c}{\Delta_{bd}} \right) \sigma_+ \right]\rho_d \nonumber \\
&\quad + \gamma_{b\downarrow} \mathcal{D} \left[ \frac{\Omega_0}{\Delta_{bd}} \left( 1 - \frac{\chi n_c}{\Delta_{bd}} \right) \left(2\sigma_+\sigma_- - \mathbf{1}\right) \right]\rho_d .
\end{align}
Similarly, for the other dissipators, we obtain the following approximations:
\begin{align}
\gamma_{b\uparrow}\mathcal{D}[\sigma_+']\rho_d &\approx \gamma_{b\uparrow} \mathcal{D} \left[ \left( 1 - \left(\frac{\Omega_0}{\Delta_{bd}}\right)^2 \left( 1 - \frac{2\chi n_c}{\Delta_{bd}} \right) \right) \sigma_+ \right]\rho_d \nonumber \\
&\quad + \gamma_{b\uparrow} \mathcal{D} \left[ \left(\frac{\Omega_0}{\Delta_{bd}}\right)^2 \left( 1 - \frac{2\chi n_c}{\Delta_{bd}} \right) \sigma_- \right]\rho_d \nonumber \\
&\quad + \gamma_{b\uparrow} \mathcal{D} \left[ \frac{\Omega_0}{\Delta_{bd}} \left( 1 - \frac{\chi n_c}{\Delta_{bd}} \right) \left( 2\sigma_+\sigma_- - \mathbf{1} \right) \right]\rho_d ,
\end{align}
\begin{align}
&\left[ \gamma_{c\downarrow} + \left(\frac{g}{\Delta_{bc}}\right)^2 \gamma_{b\downarrow} \right] \mathcal{D}[c']\rho_d \nonumber \\
&\approx \left[ \gamma_{c\downarrow} + \left(\frac{g}{\Delta_{bc}}\right)^2 \gamma_{b\downarrow} \right] \mathcal{D}[c]\rho_d \nonumber \\
&\quad + \left[ \gamma_{c\downarrow} + \left(\frac{g}{\Delta_{bc}}\right)^2 \gamma_{b\downarrow} \right] \mathcal{D} \left[ \frac{\chi\Omega_0}{\Delta_{bd}^2} \sigma_+ c \right]\rho_d \nonumber \\
&\quad + \left[ \gamma_{c\downarrow} + \left(\frac{g}{\Delta_{bc}}\right)^2 \gamma_{b\downarrow} \right] \mathcal{D} \left[ \frac{\chi\Omega_0}{\Delta_{bd}^2} \sigma_- c \right]\rho_d ,
\end{align}
\begin{align}
&\left[ \gamma_{c\uparrow} + \left(\frac{g}{\Delta_{bc}}\right)^2 \gamma_{b\uparrow} \right] \mathcal{D}[c'^\dagger]\rho_d \nonumber \\
&\approx \left[ \gamma_{c\uparrow} + \left(\frac{g}{\Delta_{bc}}\right)^2 \gamma_{b\uparrow} \right] \mathcal{D}[c^\dagger]\rho_d \nonumber \\
&\quad + \left[ \gamma_{c\uparrow} + \left(\frac{g}{\Delta_{bc}}\right)^2 \gamma_{b\uparrow} \right] \mathcal{D} \left[ \frac{\chi\Omega_0}{\Delta_{bd}^2} \sigma_- c^\dagger \right]\rho_d \nonumber \\
&\quad + \left[ \gamma_{c\uparrow} + \left(\frac{g}{\Delta_{bc}}\right)^2 \gamma_{b\uparrow} \right] \mathcal{D} \left[ \frac{\chi\Omega_0}{\Delta_{bd}^2} \sigma_+ c^\dagger \right]\rho_d .
\end{align}
Some terms can be dropped under the following approximations. The rates $\gamma_{k\uparrow}$, for $k \in \{b, c\}$, are determined by the zero-temperature decay rate $\gamma_{k,0}$ and the thermal occupation number $n_{\mathrm{th},k}$ of the environmental bath, such that $\gamma_{k\uparrow} = \gamma_{k,0}n_{\mathrm{th},k}$. Generally, the thermal population is small, so the dissipators that depend on both thermal excitation and the drive are higher-order contributions and can be ignored here. Meanwhile, for each dissipator, we keep only the leading-order drive contribution. These approximations lead to the simplified master equation
\begin{align}
\dot{\rho}_{d} &= -i \left[ H_d + H_{\mathrm{noise},d}, \rho_d \right] \nonumber \\
&\quad + \left[ \gamma_{c\downarrow} + \left(\frac{g}{\Delta_{bc}}\right)^2 \gamma_{b\downarrow} \right] \mathcal{D}[c]\rho_d \nonumber \\
&\quad + \gamma_{b\downarrow} \left[ 1 - 2\left(\frac{\Omega_0}{\Delta_{bd}}\right)^2 \right] \mathcal{D}[\sigma_-]\rho_d \nonumber \\
&\quad + 4\gamma_{b\downarrow} \left(\frac{\Omega_0}{\Delta_{bd}}\right)^2 \mathcal{D}[\sigma_+\sigma_-]\rho_d \nonumber \\
&\quad + \left[ \gamma_{c\uparrow} + \left(\frac{g}{\Delta_{bc}}\right)^2 \gamma_{b\uparrow} \right] \mathcal{D}[c^\dagger]\rho_d \nonumber \\
&\quad + \left[ \gamma_{b\uparrow} + \gamma_{b\downarrow} \left(\frac{\Omega_0}{\Delta_{bd}}\right)^4 \right] \mathcal{D}[\sigma_+]\rho_d + O\bigg(\frac{\chi^2 \Omega_0^2}{\Delta_{bd}^4}\bigg) ,
\end{align}
with
\begin{align}
H_{\mathrm{noise},d} &\approx \delta\omega_b(t) \Bigg\{ -\frac{\Omega_0}{\Delta_{bd}}\sigma_x + \left[ 1 - 2\left(\frac{\Omega_0}{\Delta_{bd}}\right)^2 \right]\sigma_+\sigma_- \nonumber \\
&\quad + \left[ \frac{g^2}{\Delta_{bc}^2} - 2\frac{\chi}{\Delta_{bd}} \left(\frac{\Omega_0}{\Delta_{bd}}\right)^2 \right]n_c \nonumber \\
&\quad + \left[ 4\frac{\chi}{\Delta_{bd}} \left(\frac{\Omega_0}{\Delta_{bd}}\right)^2 - \frac{4Kg^2}{\Delta_{bc}^3} \right]\sigma_+\sigma_- n_c \Bigg\}.
\end{align}

\section{Details of Numerical Dephasing-Rate Calculations}
\label{app:numerical_dephasing_rates}

This appendix describes the numerical procedure used to obtain the dephasing rates in Sec.~\ref{subsec:numerical_total_dephasing_rate}. We simulate the dynamics using the full Hamiltonian $H_{\mathrm{full}}(\delta\Phi)$ in Eq.~\eqref{Hfull}. The density matrix is evolved under the Lindblad master equation
\begin{align}
\dot{\rho}
&= -i\left[H_{\mathrm{full}}\big(\delta\Phi(t)\big), \rho\right]
+ \gamma_{b\downarrow}\mathcal{D}[b]\rho.
\label{eq:numerical_lindblad_dephasing}
\end{align}
where $\mathcal{D}[b]\rho=b\rho b^\dagger-\tfrac{1}{2}\{b^\dagger b,\rho\}$. The system parameters used here are the same as those in App.~\ref{app:numerics}. The FTT relaxation rate is $\gamma_{b\downarrow}=1/(20~\mu\mathrm{s})$. For simplicity, we set the thermal excitation rates of the FTT and cavity to zero. The intrinsic cavity relaxation rate is neglected here because, for our choice of system parameters, the inverse Purcell effect limits the cavity decay. In the following, we show how we generate the required $1/f$ flux noise and extract the cavity pure-dephasing time.

\begin{figure}[!htbp]
\centering
\includegraphics[width=\linewidth]{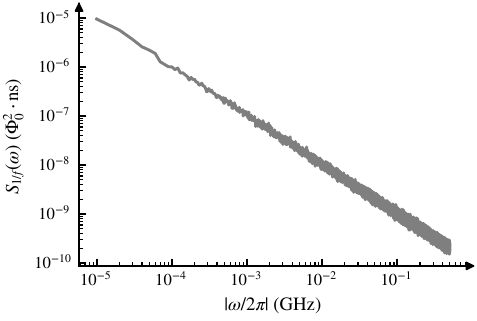}
\caption{Power spectral density of $\delta\Phi(t)$ for 100 representative $1/f$ noise trajectories generated with an infrared cutoff $\omega_{\min}/2\pi = 10~\mathrm{kHz}$ and an ultraviolet cutoff $\omega_{\max} = 1~\mathrm{GHz}$.}
\label{fig:noise_spectrum}
\end{figure}

\paragraph{Generation of $1/f$ noise.} Flux noise is incorporated through stochastic trajectories $\delta\Phi(t)$ whose power spectral densities follow a $1/f$ form. For each trajectory, Eq.~\eqref{eq:numerical_lindblad_dephasing} is solved to obtain $\rho(t)$, and the results are averaged over the flux trajectories. These trajectories are generated using the frequency-domain synthesis method from Ref.~\cite{timmer1995generatingcolornoise}. Fourier components are assigned random phases and Gaussian-distributed amplitudes with variances proportional to the target $1/f$ spectrum. An inverse Fourier transform is then applied to obtain $\delta\Phi(t)$ in the time domain. Figure~\ref{fig:noise_spectrum} shows the resulting power spectral densities for 100 representative trajectories. This ensemble size is sufficient for the quantities of interest to converge, as indicated by the small error bars in the Fig.\ \ref{fig:total_rate} (c).

\paragraph{Extraction of the dephasing time.} The states discussed in the main text as exhibiting enhanced dephasing serve as proxies for the corresponding Floquet states, as shown in App.~\ref{app:numerics}. In the numerical calculations, Floquet modes are labeled by their correspondence to bare states, as described in App.~\ref{app:numerics}. The two Floquet states used for the cavity-dephasing analysis in Sec.~\ref{subsec:numerical_total_dephasing_rate} are denoted $\ket{\phi_{g,0}(t)}$ and $\ket{\phi_{g,1}(t)}$.
Relaxation and dephasing are extracted from time evolutions initialized in two states. Relaxation is characterized by preparing
\begin{equation}
\rho(0)=\ket{\phi_{g,1}(0)}\bra{\phi_{g,1}(0)}
\end{equation}
and calculating $\mathrm{Tr}\left[\rho(t)P_{g,0}(t)\right]$, where $P_{g,0}(t)=\ket{\phi_{g,0}(t)}\bra{\phi_{g,0}(t)}$. Dephasing is characterized by preparing the equal superposition
\begin{equation}
\rho(0)=\ket{+}\bra{+},\qquad
\ket{+}=\frac{1}{\sqrt{2}}\Big(\ket{\phi_{g,0}(0)}+\ket{\phi_{g,1}(0)}\Big),
\end{equation}
and tracking the decay of $\mathrm{Tr}\left[\rho(t)X(t)\right]$, where $X(t)=\ket{\phi_{g,1}(t)}\bra{\phi_{g,0}(t)}+\mathrm{h.c.}$

The relaxation and dephasing times are obtained by fitting ensemble-averaged quantities. Specifically, $T_1$ is extracted from the rise in the ground-state population by fitting $\mathrm{Tr}\left[P_{g,0}(t)\overline{\rho}(t)\right]$ to $1-e^{-t/T_1}$, where $\overline{\rho}(t)$ denotes the density matrix averaged over different flux-noise trajectories. 

Although $1/f$ flux noise generally produces non-Markovian and nonexponential
coherence decay, the leading sensitivity to low-frequency flux fluctuations is
suppressed at the dynamical sweet spot. The remaining dephasing over the
fitting interval is therefore dominated by approximately Markovian channels
and is well described by a single exponential. Consistent with this
interpretation, the simulated coherence exhibits an approximately linear
initial decay, as expected from the short-time expansion
$e^{-t/T_2}\approx 1-t/T_2$, as discussed in Sec.~\ref{sec3}. The
pure-dephasing time then follows from
\begin{align}
\frac{1}{T_\phi}
=
\frac{1}{T_2}-\frac{1}{2T_1},
\end{align}
and we report the total dephasing rate as $\Gamma_{\phi,\mathrm{total}}=1/T_\phi$.
The data in Fig.~\ref{fig:noise_spectrum} use the same system parameters and
level truncations as those in App.~\ref{app:numerics}. Unless otherwise
stated, the fitting windows are $1~\mu\mathrm{s}$ for $T_1$ and
$5~\mu\mathrm{s}$ for $T_2$; extending these windows by doubling the durations
does not change the extracted rates. This robustness is expected: a residual
second-order sensitivity to the low-frequency $1/f$ tail would instead
manifest as a quadratic (zero-slope) onset of the coherence decay, whereas the
linear onset already noted above indicates that this channel remains
subdominant over the fitting interval; the extracted rate is therefore set by
the Markovian channels captured within a $5~\mu\mathrm{s}$ window, with
negligible drift from the low-frequency tail.

\section{Adiabatic Transitions}
\label{Adiabatic transition}
We consider adiabatic transitions between instantaneous eigenstates of the Hamiltonian
\begin{align}
H_R(t) &= \Delta_{bd} b^\dagger b + \frac{K}{2} b^{\dagger 2} b^{2} + \overline{\omega}_c c^\dagger c \nonumber \\
&\quad + \chi b^\dagger b c^\dagger c + \Omega_I(t)(b + b^\dagger).
\end{align}
In the adiabatic frame, we introduce a time-dependent unitary transformation $U_d(\Omega_I)$ that diagonalizes the instantaneous Hamiltonian $H_R(\Omega_I)$. The corresponding effective Hamiltonian is
\begin{align}
H_{\mathrm{eff}} &= U_d^\dagger(\Omega_I) \left( H_R(\Omega_I)-i\frac{\partial}{\partial t} \right) U_d(\Omega_I) \nonumber \\
&= H_d(\Omega_I)+V(\Omega_I),
\end{align}
with $V(\Omega_I)=-i U_d^\dagger(\Omega_I)\dot{U}_d(\Omega_I)$.
In the weak-drive, off-resonant regime, the diagonal contribution is well approximated by
\begin{align}
H_d(\Omega_I) &\approx \left( \Delta_{bd} + \frac{2\Omega_I^2}{\Delta_{bd}} \right)\sigma^+ \sigma^-+ \left( \tilde{\omega}_c + \frac{\Omega_I^2 \chi}{\Delta_{bd}^2} \right) c^\dagger c \nonumber \\
&\quad + \chi \left( 1 - 2 \frac{\Omega_I^2}{\Delta_{bd}^2} \right) \sigma^+ \sigma^- c^\dagger c .
\end{align}
Nonadiabatic leakage arises from the off-diagonal component of $V(\Omega_I)$. Following the calculation in Ref.~\cite{j8c7-v2hdsta}, we find that the corresponding nonadiabatic Hamiltonian in our case is $\frac{\dot{\Omega}_I}{\Delta_{bd}}\sigma_y$.
Suppressing leakage requires the nonadiabatic coupling to remain small compared with the instantaneous gap, which is of order $|\Delta_{bd}|$:
\[
\left|\frac{\dot{\Omega}_I}{\Delta_{bd}}\right| \ll |\Delta_{bd}|.
\]
For a ramp with characteristic amplitude $\Omega_0$ applied over a ramp time $\tau$, so that $|\dot{\Omega}_I| \approx \Omega_0/\tau$, this condition implies the lower bound
\begin{equation}
\tau \gg \frac{\Omega_0}{|\Delta_{bd}|^2}
\overset{\eqref{eq:ss_unselective}}{=}
\sqrt{\frac{1}{4K\Delta_{bd}}}.
\end{equation}
Increasing $|\Delta_{bd}|$ can reduce the required $\tau$ for preparing the desired state.

\begin{figure}[!htbp]
\centering
\includegraphics[width=\linewidth]{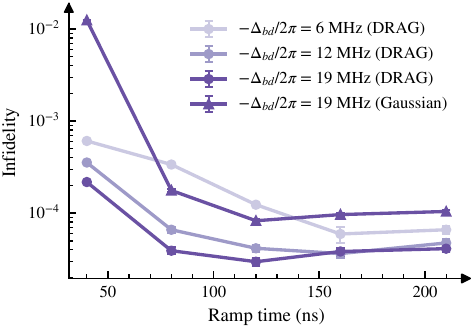}
\caption{State-preparation infidelity $\mathcal{I}(\tau)$ as a function of ramp time $\tau$ for several detunings $\Delta_{bd}$, comparing Gaussian ramps with DRAG ramps. The nonmonotonic dependence reflects the trade-off between coherent leakage from unwanted transitions, which is suppressed for longer $\tau$, and decoherence, which accumulates for longer $\tau$.}
\label{fig:ramp}
\end{figure}

To bypass this speed limit, we implement a shortcut to adiabaticity~\cite{j8c7-v2hdsta} using a DRAG-style control field $\Omega_Q$.
Specifically, we use the total drive Hamiltonian
\begin{equation}
H_{\text{total}} = H_R - i \Omega_Q (b - b^\dagger),
\end{equation}
and choose the quadrature amplitude to cancel the nonadiabatic coupling:
\begin{equation}
\Omega_Q = -\frac{\dot{\Omega}_I}{\Delta_{bd}}.
\end{equation}
The effective Hamiltonian in the adiabatic frame then contains no leakage term, so the evolution remains diagonal in the instantaneous eigenbasis even for fast ramps.

Numerical simulations confirm the qualitative analysis above. We compute the state-preparation infidelity at the end of the ramp as a function of the ramp time $\tau$ for several detunings $\Delta_{bd}$.
The infidelity is defined as
\begin{align}
\label{eq:infidelity_def}
\mathcal{I}(\tau)
&= 1-\frac{1}{2}\big[ \mathrm{Tr}\left(P_{g0}(\tau)\bar{\mathcal{E}}_{\tau}\left(\rho_{0,g0}\right)\right) \nonumber \\
&\quad + \mathrm{Tr}\left(P_{g1}(\tau)\bar{\mathcal{E}}_{\tau}\left(\rho_{0,g1}\right)\right)\big],
\end{align}
where $P_{g0}(\tau)=\ket{\phi_{g0}(\tau)}\bra{\phi_{g0}(\tau)}$ and
$P_{g1}(\tau)=\ket{\phi_{g1}(\tau)}\bra{\phi_{g1}(\tau)}$ project onto the desired Floquet states at the end of the pulse. The map
$\bar{\mathcal{E}}_{\tau}(\cdot)=\langle \mathcal{E}_{\tau}^{\delta\Phi}(\cdot)\rangle_{\delta\Phi}$
denotes the evolution averaged over flux-noise trajectories $\delta\Phi(t)$, with the evolution for each trajectory obtained from the master equation in Eq.~\eqref{eq:numerical_lindblad_dephasing}.
The initial states are
$\rho_{0,g0}=\ket{\bar{g}0}\bra{\bar{g}0}$ and
$\rho_{0,g1}=\ket{\bar{g}1}\bra{\bar{g}1}$, where the overbar denotes eigenstates of the static system Hamiltonian in the dispersive regime.
The results are shown in Fig.~\ref{fig:ramp}. For a fixed detuning, the infidelity initially decreases with increasing $\tau$ because a slower ramp improves adiabaticity and suppresses coherent leakage, but it eventually increases as the longer evolution accumulates more decoherence-induced error. For a fixed ramp time, for example, $\tau=50~\mathrm{ns}$, a larger detuning yields lower infidelity because of the larger instantaneous gap. Finally, for the same detuning and ramp time, the DRAG ramp outperforms the Gaussian ramp because the added $Q$-quadrature control enhances adiabaticity by suppressing unwanted transitions.

\section{System Parameters Used in the Simulations}
\label{System Parameters Used in the simulations}
For the case of binomial encoding, we use the system parameters given in App.~\ref{app:numerical_dephasing_rates}. The only difference is that we choose $g/2\pi=25~\mathrm{MHz}$ and truncate cavity to $5$ levels.

For the case of dual-rail encoding, we choose the following parameters: $g_1/2\pi=g_2/2\pi=50~\mathrm{MHz}$, $\omega_{c}^{(1)}/2\pi = 5.226~\mathrm{GHz}$, $\omega_{c}^{(2)}/2\pi = 7.335~\mathrm{GHz}$,
$\omega_{c}^{(3)}/2\pi = 4.292~\mathrm{GHz}$,
and $\omega_{c}^{(4)}/2\pi = 8.458~\mathrm{GHz}$. The FTT parameters are the same as those given in App.~\ref{app:numerical_dephasing_rates}.
\bibliography{ref}
\end{document}